\documentclass{article}
\usepackage{arxiv}

\makeatletter
\def\ps@pprintTitle{%
 \let\@oddhead\@empty
 \let\@evenhead\@empty
 \def\@oddfoot{}%
 \let\@evenfoot\@oddfoot}
\makeatother
\usepackage[utf8]{inputenc}
\usepackage{lineno,hyperref}
\usepackage{graphicx}
\usepackage{multirow}
\usepackage[table,xcdraw]{xcolor}
\usepackage{amsmath}
\usepackage{amsfonts}
\usepackage[nolist]{acronym}
\usepackage{soul}
\usepackage{arydshln}
\usepackage{subcaption}
\usepackage{comment}
\usepackage{geometry}
\usepackage[super]{nth}
\usepackage{pgfplots}
\pgfplotsset{compat=newest}
\usepgfplotslibrary{external}
\usepgfplotslibrary{patchplots}
\usetikzlibrary{shapes,arrows}
\usetikzlibrary{positioning,calc}
\usetikzlibrary{decorations.pathreplacing,decorations.markings,math,arrows.meta}
\usetikzlibrary{spy}
\usetikzlibrary{pgfplots.groupplots}
\usepgfplotslibrary{fillbetween}
\usetikzlibrary{patterns}
\newbox{\bigpicturebox}

\pgfdeclarelayer{ft}
\pgfdeclarelayer{bg}
\pgfsetlayers{bg,main,ft}

\DeclareMathOperator{\E}{\mathbb{E}}
\def\equationautorefname~#1\null{Equation~(#1)\null}

\newcommand{\bX}{\mbox{\boldmath$X$}}
\newcommand{\bR}{\mbox{\boldmath$R$}}
\newcommand{\bF}{\mbox{\boldmath$F$}}
\newcommand{\bL}{\mbox{\boldmath$L$}}
\newcommand{\bSigma}{\boldsymbol{\Sigma}}
\newcommand{\bZ}{\mbox{\boldmath$Z$}}
\newcommand{\bW}{\mbox{\boldmath$w$}}
\newcommand{\bS}{\mbox{\boldmath$S$}}
\newcommand{\bV}{\mbox{\boldmath$V$}}
\newcommand{\bM}{\mbox{\boldmath$M$}}
\newcommand{\bg}{\mbox{\boldmath$g$}}
\newcommand{\bu}{\mbox{\boldmath$u$}}
\newcommand{\bU}{\mbox{\boldmath$U$}}
\newcommand{\bq}{\mbox{\boldmath$q$}}
\newcommand{\bp}{\mbox{\boldmath$p$}}
\newcommand{\bmu}{\mbox{\boldmath$\mu$}}

\newcommand\figureFontSize{9}

\begin{acronym}
\acro{MOR}{Model Order Reduction}
\acro{RB}{Reduced Basis}
\acro{FOM}{Full Order Model}
\acro{ROM}{Reduced Order Model}
\acro{ROMs}{Reduced Order Models}
\acro{GAN}{Generative adversarial network}
\acro{SHM}{Structural Health Monitoring}
\acro{FE}{Finite Element}
\acro{DOF}{degree of freedom}
\acro{POD}{Proper Orthogonal Decomposition}
\acro{PCA}{Principal Component Analysis}
\acro{SVD}{Singular Value Decomposition}
\acro{VAE}{Variational AutoEncoder}
\acro{cVAE}{conditional Variational AutoEncoder}
\acro{LHS}{Latin Hypercube Sampling}
\acro{NARX}{Nonlinear AutoRegressive with eXogenous input}
\end{acronym}

\title{A Reduced Order Model conditioned on monitoring features for estimation and uncertainty quantification in engineered systems}

\author{Konstantinos Vlachas \\
    Dept. of Civil, Environmental and Geomatic Engr.\\
    ETH Zurich\\
    Zurich, Switzerland\\
    \texttt{vlachas@ibk.baug.ethz.ch} 
    \And
    Thomas Simpson \\
    Dept. of Civil, Environmental and Geomatic Engr.\\
    ETH Zurich\\
    Zurich, Switzerland\\
    \texttt{tom.simpson74@gmail.com}\\
    \And
    Anthony Garland \\
    Component Sciences and Mechanics\\
    Sandia National Laboratories\\
    Albuquerque, New Mexico, USA, \\
    \texttt{agarlan@sandia.gov}
    \And
    D. Dane Quinn \\
    Department of Mechanical Engineering\\
    The University of Akron\\
    Akron, Ohio, USA\\
    \texttt{quinn@uakron.edu}
    \And
    Charbel Farhat \\
    Department of Aeronautics and Astronautics\\
    Stanford University\\
    Stanford, California, USA\\
    \texttt{cfarhat@stanford.edu}
    \And
    Eleni Chatzi \\
    Dept. of Civil, Environmental and Geomatic Engr.\\
    ETH Zurich\\
    Zurich, Switzerland\\
    \texttt{chatzi@ibk.baug.ethz.ch}
}

\begin{document}
\maketitle

\begin{abstract}
Reduced Order Models (ROMs) form essential tools across engineering domains by virtue of their function as surrogates for computationally intensive digital twinning simulators. 
Although purely data-driven methods are available for ROM construction, schemes that allow to retain a portion of the physics tend to enhance the interpretability and generalization of ROMs.
However, physics-based techniques can adversely scale when dealing with nonlinear systems that feature parametric dependencies.
This study introduces a generative physics-based ROM that is suited for nonlinear systems with parametric dependencies and is additionally able to quantify the confidence associated with the respective estimates.
A main contribution of this work is the conditioning of these parametric ROMs to features that can be derived from monitoring measurements, feasibly in an online fashion. 
This is contrary to most existing ROM schemes, which remain restricted to the prescription of the physics-based, and usually a priori unknown, system parameters. 
Our work utilizes conditional Variational Autoencoders to continuously map the required reduction bases to a feature vector extracted from limited output measurements, while additionally allowing for a probabilistic assessment of the ROM-estimated Quantities of Interest. 
An auxiliary task using a neural network-based parametrization of suitable probability distributions is introduced to re-establish the link with physical model parameters.
We verify the proposed scheme on a series of simulated case studies incorporating effects of geometric and material nonlinearity under parametric dependencies related to system properties and input load characteristics.
\end{abstract}

\keywords{parametric reduction \and  Reduced Order Models (ROMs) \and  conditional VAEs \and  uncertainty quantification}

\section{Introduction} \label{Introduction}

Reduced order modeling techniques can be classified into two main categories: \emph{purely data-driven} \cite{montans2019data,peherstorfer2016data} approaches and physics-based methods \cite{benner2017model,chinesta2014pgd}.
\emph{Purely data-driven} frameworks typically learn the dynamic behavior through input-output relationships from simulated data \cite{bacsa2023symplectic} or capitalize on the availability of monitoring information from the system of interest \cite{brunton2022data, kivcic2023adaptive}.
The continuous advancement of computational power and the increasingly sophisticated (deep) machine learning architectures have enhanced the capability of these methods to more closely approximate intricate dynamics, even in complex scenarios \cite{moya2022digital,najera2024structure, nikolopoulos2024ai}.
However, the efficacy of such methods is contingent upon the scope and quality of the data used for training \cite{duthe2023graph,fresca2023long,vlachas2018data}.
In contrast, \emph{physics-based} models attempt to incorporate the underlying principles of physics (e.g. mechanics or dynamics); a task which can be achieved with various degrees of prescription \cite{haywood2023discussing,von2021informed}.
A particular subset of such \emph{physics-based} \ac{ROMs}, reflecting a heavier degree of physics-prescription corresponds to projection-based \ac{ROMs} \cite{agathos2024accelerating}.
These are typically derived from a dynamical system's equations of motion using suitable techniques relying on \ac{RB} schemes \cite{amsallem2009method,hesthaven2022reduced,qian2020lift} and are thus characterized as intrusive techniques. 
Alternative non-intrusive methodologies that do not require the assembly of a \ac{RB} have also been proposed \cite{mahdiabadi2021non,wu2019modal}, with the most notable example being the Proper Generalized Decomposition \cite{chinesta2011short,pasquale2024modular}.

Regardless of the specific modeling category, the reduction task is particularly challenged when parametric dependence is involved, particularly when tasked with representing a system that exhibits significantly differentiated dynamics for varying system properties or under diverse rages of excitation \cite{cicirello2022machine,najera2023structure}.
The reduction task is further challenged by the requirements in terms of compute time, especially when complex and detailed \ac{FOM} models are considered and the derived \ac{ROMs} are required to perform (near) real-time estimation while providing confidence for the respective predictions \cite{pasquale2021separated,tezaur2022robust}.
In this work, we choose to tackle the reduction of large-scale nonlinear parametric systems through the use of a projection-based scheme, making use of \ac{POD} \cite{amsallem2012nonlinear,Kerschen2005}. 
Since the \ac{POD} is a linear operator and only provides a \emph{linear} subspace approximate to the actual response manifold \cite{gobat2022reduced}, constructing a single global \ac{POD} basis for nonlinear parametric systems can become inefficient or even fail \cite{antoulas2020interpolatory}.
To this end, the \ac{POD}-based projection is typically applied locally, forming a pool or directory of \ac{POD} bases, thus being able to capture local dynamics \cite{amsallem2016pebl, berthelin2022disciplinary}.

In this context, \ac{POD} can be applied locally with respect to time, thus deriving reduction bases corresponding to a certain time window of the simulation \cite{mlinaric2023unifying,amsallem2012nonlinear}. 
Alternatively, each of the local \ac{ROMs} might correspond to a certain region of the parameter space \cite{morsy2022reduced, he2020situ, haasdonk2011training}.
In both cases, clustering or interpolation strategies between the library of local candidates are required during the low-order model evaluations \cite{amsallem2008interpolation,ghavamian2017pod}.
The interested reader is referred to \cite{quarteroni2014reduced} for a more detailed discussion on local \ac{ROMs}, whereas stability considerations are discussed in \cite{friderikos2022stability}.

While such techniques can be viewed as well-established for generating an actionable \ac{ROM} for \ac{SHM} applications, their effectiveness is heavily contingent on the method employed for selecting the proper basis from the pre-assembled library \cite{goutaudier2023new, touze2021model}.
On the one hand, approaches that utilize interpolation or clustering with suitable physics-inspired measures might still suffer from robustness or generalization issues when addressing highly nonlinear dynamics \cite{agathos2020parametrized}.
For example, if the basis interpolation is based on linear operations or the local \ac{ROM}selection scheme relies on naive classifiers, it is often common to anticipate larger errors in the approximation for specific regions of the parameter space. 
Greedy or adaptive sampling during training equipped with suitable error indicator can be used to account for this shortcoming, however, such a process elevates complexity and may require a nontrivial training and implementation process \cite{paul2015adaptive, rozza2013reduced}. 

In addressing these challenges, we here leverage the benefits of both \emph{purely data-driven} and \emph{physics-based} approaches, drawing inspiration from recent advancements in \ac{SHM} that utilize machine- or deep-learning algorithms to enhance the effectiveness and performance of \ac{ROMs} while achieving an automated training process \cite{angeli2021deep,cicci2022deep, lai2022neural}.
The core of the framework in this work is built upon the recent contribution in \cite{simpson2023vprom}, which proposes a generative model as an approximator of the generalized mapping between parametric inputs and local projection bases while maintaining a projection-based \ac{ROM} that guarantees robustness and accuracy. 
Principally, a generative model learns the joint distribution $P(X,Y)$ of the observed data $X$ and labels $Y$ \cite{girin2020dynamical,harshvardhan2020comprehensive}.
Deep generative models have become a popular approach to forming such generative models that are flexibly trained and are suitable for capturing complex nonlinear system dependencies \cite{ruthotto2021introduction}.
A primary example is found in the \ac{VAE} \cite{Kingma2014} model that has enjoyed broad adoption across diverse fields \cite{hawkins2021generating,lopez2020enhancing}.

In the context of model order reduction for engineering, such probabilistic models are typically associated with \emph{data-driven} approaches \cite{romor2023non}.
The work of \cite{lee2020model}, adopts a \ac{VAE} to approximate the nonlinear response manifold for model reduction and has resulted in numerous follow-ups.
These include works in transonic flow reconstruction \cite{kang2022physics}, unsteady flow modeling \cite{eivazi2020deep}, as well as fatigue estimation in wind turbine blades \cite{Mylonas2021} and nonlinear dynamics problems \cite{conti2023reduced,Simpson2021,vlachas2022multiscale}. 
The comprehensive review offered in \cite{szalai2023data}, reports on the ability of \ac{VAE}-based strategies to learn highly nonlinear response manifolds and address high dimensional input-output data.

We here adopt such a nonlinear generative modeling approach, and specifically opt for utilization of a \ac{cVAE}, to derive local projection bases, and thus local \ac{ROM}s, at unseen system parameter values.
The proposed model, termed \emph{VpROM}, essentially learns the joint distribution of system parameters and associated local bases, offering a flexible means to generate the local \ac{ROM} basis for any given parameter sample.
Essentially, the \ac{VAE} component is used to perform interpolation in the inferred latent space, the smoothness of which ensures superior performance for basis generation \cite{amsallem2016pebl, kapteyn2022data}. 
Moreover, the probabilistic nature of the \ac{VAE} allows for an indirect uncertainty quantification in the delivered response estimate when performing this inference task.

The derived \ac{ROM} reflects a coupling of \emph{data-driven} and \emph{physics-based} techniques.
Compared to the aforementioned approaches employing generative models in model order reduction, e.g., \cite{fresca2022pod,lee2020model,Simpson2021}, our framework does not necessitate an additional machine-learning network to learn the propagation of dynamics forward in time on the latent space; rather, it retains the projection-based reduction, thus integrating the dynamics using the equation of motion in a suitable low-order coordinate set. 
The resulting \ac{ROM} provides a low-order, yet still physics-based, representation of the physical space of the model and can reproduce the \ac{FOM} response at any specified physical field of interest.

Furthermore, our work, although relying on the prior framework presented in \cite{simpson2023vprom} as mentioned, is differentiated from this via two distinctive extension features: 
First, a primary novelty lies in linking the physical parameter space of the derived \ac{ROM} to monitoring features that can be automatically inferred from limited response measurements.
Within the \ac{SHM} context, this offers a more practical conditioning, for the selection of an appropriate \ac{ROM} basis, as opposed to relying on physical system parameters, which are typically unknown a priori.
To achieve this, the \ac{cVAE} is conditioned to response features that are extracted from the acceleration signals and not to the physical system parameters.
Then, an auxiliary task - using neural networks - is introduced to re-establish the required connection between the monitoring features used for conditioning and the parameter values, which are required by the derived \ac{ROM} to propagate the dynamics in the reduced space. 
Second, the training of the \ac{cVAE} is improved to address generalization and robustness issues. 
Specifically, the loss term of the \ac{cVAE} is modified during the latter training epochs to include the reconstruction error of the local projection bases and the precision of the \ac{ROM} estimation with respect to the \ac{FOM} response. 

The structure of this paper is organized as follows:
\autoref{sec:ReducedOrderModelling} offers the background knowledge on projection-based reduced order modeling and the methodological elements of our work for treating nonlinear parametric systems. 
Then, \autoref{sec:VpROM} describes the \ac{cVAE} model utilized in place of current state-of-the-art interpolation or clustering methods in parametric \ac{ROM}s.
The framework's training and prediction mode is reported in detail, along with the architecture of the neural network that was deployed to recover the physical parameters that correspond to the monitoring features that are inferred from sensing data. 
In \autoref{sec:Results}, the proposed approach is demonstrated in two synthetic (simulated) case studies. 
Firstly, a two-dimensional arc structure with geometric nonlinearities due to large deformations is modeled, with parametric behavior depending on system properties.
Secondly, the approach is demonstrated on a large-scale, three-dimensional \ac{FE} model of a welded kingpin connection undergoing plastic deformation, depending on both system and excitation characteristics.
In both cases, the proposed \ac{cVAE} framework provides reliable confidence bounds that capture the frequency trends and amplitude characteristics of the underlying response, proving to yield a robust response predictive \ac{ROM}, which is conditioned on monitoring features that are inferable in an online manner.
Finally, \autoref{sec:Conclusions} concludes the work by summarising the numerical results achieved and the limitations of the framework, offering, in turn, perspectives on future developments.

\section{Physics-based Reduced Order Modelling} \label{sec:ReducedOrderModelling}

Our work applies physics-based reduction to parameterized dynamical systems, resulting in a low-order surrogate of a \ac{FOM}.
The derived representation, termed \ac{ROM}, provides accelerated system evaluations, which are useful for downstream tasks in \ac{SHM} diagnostics and decision support for the operation and maintenance planning of engineered systems.
In what follows, the problem formulation is introduced first in the form of the governing nonlinear equations of motion. 
Then, the projection-based \ac{ROM} is described, followed by the framework components needed to address parametric dependencies.
Last, the computational efficiency bottleneck when propagating the response in the low-order subspace is discussed.

\subsection{Problem Statement} \label{ProblemStatement}
The \ac{FOM} assumed in this work corresponds to a \ac{FE} model that is suitable for nonlinear structural dynamics simulations. 
The corresponding nonlinear dynamical system is dependent on the input vector $\bp=[p_1,...,p_{\mathrm{k}}]^{\mathrm{T}} \in \Omega \subset \bR^k$, which captures all system- and excitation-relevant parameters. 
In turn, the response of such a system is described by the following set of governing equations:
\begin{equation}
\bM\ddot{\bu}(t) + \bg\left(\bu(t), \dot{\bu}(t), \bp \right) = \bF (t,\bp),
\label{eq:FOM}
\end{equation}
\noindent
where $\bu(t) \in \bR^n$ represents the system's behavior in terms of displacements, $\bM \in \bR^{n \times n}$ denotes the mass matrix, and  $\bF(t, \bp) \in \bR^{n}$ the induced excitation.
For the sake of simplicity, the parametric dependency in the mass matrix has been dropped.
The nonlinearities are captured via the restoring force term $\bg\left(\bu(t), \dot{\bu}(t), \bp \right) \in\bR^{n}$ that potentially models complex nonlinear effects, ranging from plasticity to hysteresis or interface nonlinearities, which, in turn, depend on the response of the system and its parameters $\bp$.
The \ac{FE} model used as the \ac{FOM} corresponds to a discretized version of \autoref{eq:FOM}.
The full-order dimension of the \ac{FOM} is expressed by variable $n$ that represents the size of the coordinate space and, thus, the total number of degrees of freedom.
This variable indirectly measures the computational complexity and resources needed to evaluate the \ac{FOM}.

\subsection{Projection-based model order reduction}
The reduction approach employed herein relies on a Galerkin projection-based scheme to derive a low-order representation for the dynamic problem as presented in \autoref{ProblemStatement}.
A more generalizable extension using the Petrov-Galerkin method is also possible, as presented in \cite{barnett2022quadratic,xiao2013non}.
As explained in \autoref{Introduction}, we have opted for a projection-based technique for the \ac{ROM} to be able to recover the response at any given physical field of interest and on the full physical space of the model at any time.
Thus, the derived \ac{ROM} is not limited to inferring specific response quantities or only at a few nodes; rather it can capture the displacement response of \autoref{eq:FOM} whilst inferring stresses, strains, and accelerations at once. 
This, in turn, increases its interpretability and its serviceability for higher-level \ac{SHM} systems \cite{tatsis2022hierarchical, di2023physics}. 

To begin with, the reduction approach employed herein relies on the availability of a full-order representation of the system, termed \ac{FOM}.
This is a \ac{FE} model that spatially discretizes the governing equations in \autoref{eq:FOM}.
In addition, projection-based techniques depend on the underlying assumption that the dynamic behavior, namely the solution of \autoref{eq:FOM}, lies in a low-order subspace, whose dimension $r$ is orders of magnitude smaller than the full-order dimension of the problem, denoted by $n$ ($r \ll n$).
Therefore, the following approximation holds:
\noindent
\begin{equation}
	\bu\left(\bp\right) \approx \bV\left( \bp \right) \bq
\label{eq:project}
\end{equation}
\noindent where $\bV \in \mathbb{R}^{N \times r}$ represents the projection-bases that expresses the aforementioned subspace of the \ac{ROM} and $\bq \in \mathbb{R}^{r}$ is the respective low-order coordinate vector.
Via substitution of $\bu$ into \autoref{eq:FOM} and after multiplying the governing set with $\bu^{T}$, thus performing a Galerkin projection, the following can be derived:
\begin{equation}
\Tilde{\bM} \ddot\bq\left( t \right) + \Tilde{\bg}\left(\ddot\bq,\dot\bq, \bp \right) = \Tilde{\bF}\left( \bp, t \right)
\label{eq:ROM}
\end{equation}
where $\Tilde{\bM}=\bV^T\bM\bV$, $\Tilde{\bg}=\bV^T\bg$ and $\Tilde{\bF}=\bV^T \bF$.
The key to an accurate \ac{ROM} able to capture the \ac{FOM} dynamics and reproduce them efficiently is the assembly of the projection basis $\bV$.
To this end, typically, the \ac{POD} technique is employed, which relies on a set of training evaluations of the \ac{FOM} of \autoref{eq:FOM} for the respective set of parameters and collects the solutions as follows:
\noindent
\begin{equation}
	\hat{\bS} = \left[
	\begin{array}{c c c c}
		\hat{\bU}\left( \bp_1 \right) & \hat{\bU}\left( \bp_2 \right) & \ldots & \hat{\bU}\left( \bp_{N_s} \right)
	\end{array}
	\right]
	\label{eq:Snaps}
\end{equation}
\noindent where $\hat{\bU}\left( \bp_i \right) \in \mathbb{R}^{N \times N_t}$ contains the displacement time history for every \ac{DOF} for a given parametric realization $\bp_i$, henceforth termed as a snapshot, and, as a result, $\hat{\bS} \in \mathbb{R}^{N \times (N_t \times N_s)}$ is termed the snapshot matrix. The variable $N_t$ represents the number of simulation time steps, $\bp_i$ is the parameter vector for snapshot $i$, and $N_s$ is the total number of snapshots.
Via \ac{SVD} of $\hat{\bS}$ the \ac{ROM} projection basis can be assembled:
\noindent
\begin{equation}
	\hat{\bS} = \bL \bSigma \bZ^T
    \label{eq:POD}
\end{equation}
\noindent and after truncating $\bL$:
\begin{equation}
	\bV = \left[
	\begin{array}{c c c c}
		\bL_1 & \bL_2 & \ldots & \bL_r
	\end{array}
	\right]
    \label{eq:Vmodes}
\end{equation}
\noindent where $\bL_i$ is the $i$ column of matrix $\bL$, termed as \ac{POD} mode.
Since the low-order dimension of \autoref{eq:project} is defined as $r$, the above truncation is applied to obtain the first $r$ orthonormal components of $\bV$.
To define $r$, a suitable error measure is employed based on the singular value decay.
More details are provided in \ref{sec:app2}.

\subsection{CpROM: Treatment of parametric dependencies via local Basis Coefficients ROMs} \label{CpROM}

The system's dynamic behavior, governed by the equations in \autoref{eq:FOM}, is dictated by the parameter vector $\bp$ representing system properties and characteristics of the external excitation.
Thus, the resulting response is highly dependent on the parameter vector realization and often showcases localized effects in the parametric space due to dedicated nonlinear terms being activated.
As a result, attempting a projection-based reduction with a single basis, as described in \autoref{eq:project}, might necessitate a large number of modes, resulting in a prohibitively large dimension and a \ac{ROM} that proves inefficient or impractical \cite{agathos2020parametrized}.
To address this, the typical strategy is to assemble a directory or library of local \ac{POD} bases $\bV_i$, each one derived using $\hat{\bU}\left( \bp_i \right)$, namely \ac{FOM} snapshots for a specific realization $\bp_i$ of the parameter vector.
This approach enables the ROM to capture localized effects while utilizing suitable interpolation or clustering techniques for approximating the responses in intermediate parameter samples \cite{amsallem2012nonlinear, antoulas2020interpolatory}.

An alternative strategy for the treatment of parametric dependencies in the context of nonlinear \ac{ROM}s has been proposed in \cite{Vlachas2021}.
Specifically, the local bases are projected first to the tangent space of the proper Grassmannian manifold, where interpolation is typically carried out on the existing state of the art approaches \cite{amsallem2008interpolation,friderikos2020space}.
This projection to the tangent space is required to ensure that any interpolation operations performed on the tangent space output a local basis that retains key properties like orthogonality. 
Then, in \cite{Vlachas2021}, a two-stage process is introduced:\\
First, the following system is solved in the least-squares sense:
\begin{equation}
    \bV_i = \bV_{global} \bX_i
    \label{eq:coeffs}
\end{equation}
\noindent
where $\bV_i \in \mathbb{R}^{n \times r}$ is an instant of the library of local bases, $\bV_{global} \in \mathbb{R}^{n \times \tilde{r}}$ captures the dynamics across the entire domain and $ \bX_i \in \mathbb{R}^{\Tilde{r} \times r}$ is the respective coefficient matrix. 
Second, interpolation (or any equivalent operation) is performed on the coefficient matrices $\bX_i$.
These comprise a reduced size ($\Tilde{r} \ll r$), thus removing any dependency on the \ac{FOM} dimension $n$, while the additional projection enables the dependence on $\bp$ to be formulated on a separate level from that of the snapshots or the local bases.
The approximation in \autoref{eq:coeffs} is carried out on the tangent space, thus the respective variables denote the projected quantities. 
For simplicity purposes, we did not introduce an additional variable and retained the same notation for the bases $\bV_i$.
For the sake of completeness, $\Tilde{r}$ signifies the total number of truncated odes retained on $\bV_{global}$ and can be computed similarly to $r$. 
Then, the respective matrices $\bX_i$ are to be interpolated properly and projected back to the original space to compute the respective local \ac{ROM} basis $\bV$ for any validation parametric sample. 
This back-and-forth projection strategy is required for the resulting subspaces to retain orthogonality and positive-definiteness required for subsequent forward integration of the \ac{ROM} in \autoref{eq:ROM} in time.
The interested reader is referred to \cite{Vlachas2021} for a more in-depth presentation.

\subsection{Hyper-reduction}\label{sec:hyperreduction}

Hyper-reduction refers to a second-tier approximation required to tackle the computational bottleneck of updating and reconstructing the \ac{ROM} system matrices efficiently \cite{peherstorfer2014localized}.
The Energy Conserving Mesh Sampling and Weighting (ECSW) technique is employed herein \cite{farhat2014dimensional}. 
ECSW selects a subset of the total elements of the spatial discretization of the \ac{FE} model, which is employed here as the \ac{FOM}, by solving a nonlinear optimization problem. 
The optimization problem is formulated in a physics-based manner, assuming that a weighted evaluation of the corresponding projections of the nonlinear terms only at the aforementioned elements' subset can accurately approximate the total (internal) work. 
After solving the optimization and defining the elements of the subset and the corresponding weights, a hyper-reduced \ac{ROM} can be derived, which provides substantial computational toll reduction.
Prior works extensively address the methodology and discuss alternative approaches, which are, therefore, outside the scope of this paper. 
The reader can refer to \cite{Farhat2015,grimberg2021mesh, peherstorfer2014localized}.

\section{VpROM: Generative modeling for local ROMs} \label{sec:VpROM}

Parametric dependencies are treated in the \ac{ROM} level using a library of local projection bases and suitable clustering or interpolation techniques described in the previous section. 
Our work utilizes a nonlinear generative model in the form of a \ac{cVAE} to substitute these methods and address robustness and performance issues by allowing a more generalizable nonlinear mapping in the parameter-basis relation to be captured.  
In addition, the derived \emph{VpROM} quantifies the confidence while performing response estimation tasks, thus offering an increased utility for \ac{SHM} applications. 

\subsection{Variational Autoencoder (VAE)}

The \ac{VAE} architecture is a generative model that assumes the input data, termed observations, can be characterized by unobserved \emph{latent} variables \cite{Kingma2014}. 
Thus, the \ac{VAE} typically utilizes deep neural networks to infer the underlying relationship between these latent variables and the observed quantities. 
Concerning model order reduction, a \ac{VAE} can be considered as an equivalent Bayesian implementation of its deterministic counterpart, which has been extensively used for dimensionality reduction \cite{kim2022fast, pichi2024graph,wu2021reduced}.

The \ac{VAE} assumes that the observations $\bX$ are characterized by a distribution $p(\bX)$ to be approximated via a simplified distribution $p_{\phi}(\bX)$.
In our work, the term observations refers to the coefficient matrix $\bX$ formulated in \autoref{eq:coeffs}, expressing the local low-order subspaces with respect to a global basis spanning the entire domain of parametric inputs. 
In addition, all bases come from synthetically generated snapshots of \ac{FOM} evaluations.
Next, we also assume that the complex underlying distribution of the observed data is driven by a low-order variable set $\bZ$, which follows a simpler distribution initially approximated with a prior $p(\bZ)$.
The goal here is to exploit a sufficiently powerful and flexible approximator, termed the \emph{decoder}, to map the latent variables $\bZ$ to the complexly distributed data $\bX$ by learning the distribution $p_{\phi}(\bX\vert \bZ)$. 
Deep neural networks are suitable for this task \cite{Doersch2016}. 
The \emph{decoder} network is additionally parameterized by $\phi$, corresponding to its weights and biases. 
Thus, the following expression holds:
\begin{equation}
    p_{\phi}(\bX) = \int{p_{\phi}(\bX\vert \bZ)p(\bZ)dz}
    \label{eq:VAE}
\end{equation}

During the training of the generative model in \autoref{eq:VAE}, the parameters $\phi$ that maximize the likelihood of the input observations are inferred. 
The respective expression is:
\begin{equation}
    \phi = \operatorname*{argmax}_\phi \prod_{i=1}^{N}\int{p_{\phi}(\bX^i\vert \bZ)p(\bZ)dz}
    \label{eq:phis}
\end{equation}

\begin{figure}[h]
    \centering
    \tikzstyle{featrec} = [rectangle, rounded corners, 
draw=cyan!60, ultra thick, minimum width=0.5cm, 
minimum height=1.25cm,
text centered]

\tikzstyle{latent} = [rectangle, rounded corners, 
draw=cyan!60, ultra thick, minimum width=0.5cm, 
minimum height=0.8cm,
text centered]

\tikzstyle{latentz} = [rectangle, rounded corners, 
draw=cyan!60, ultra thick, minimum width=0.5cm, 
minimum height=1.0cm,
text centered]

\tikzstyle{texting} = [rectangle, text centered]

\tikzstyle{coeffrec} = [rectangle, rounded corners, 
draw=cyan!60, ultra thick, minimum width=0.5cm, 
minimum height=3.75cm,
text centered]

\tikzstyle{kiklos} = [circle,
draw=black, thick,  minimum size = 0.3cm,inner sep=0pt,
text centered]

\tikzstyle{enc} = [trapezium, shape border rotate=270, rounded corners, 
draw=cyan!60, ultra thick, trapezium stretches=false, 
trapezium left angle=45, 
trapezium right angle=45, 
minimum width=1cm, 
minimum height=1cm, text centered]

\tikzstyle{dec} = [trapezium, shape border rotate=90, rounded corners, 
draw=cyan!60, ultra thick, trapezium stretches=false, 
trapezium left angle=45, 
trapezium right angle=45, 
minimum width=1cm, 
minimum height=1cm, text centered]

	\begin{tikzpicture}
		\node (coeffs) [coeffrec] {{$\bX$}};
		\node (encoder) [enc, align=center, right of=coeffs, xshift=0.75cm] {Encoder \\ $f_{\theta}\left( \bX \right)$};
		\draw[-Stealth, thick] (coeffs.east) -- (encoder.west);
		\node (mu) [latent, right of=encoder, xshift=0.75cm, yshift=0.75cm] {$\mu$};
		\node (sigma) [latent, below of=mu, yshift=-0.5cm] {$\sigma$};
		\node (epsilon) [latent, below of=sigma] {$\varepsilon$};
		\draw[-Stealth, thick] (encoder.east) -- (mu.west);
		\draw[-Stealth, thick] (encoder.east) -- (sigma.west);
		\node (kik2) [kiklos, right of=sigma, xshift=-0.2cm, yshift=-0.5cm] {$\odot$};
		\node (kik3) [kiklos, right of=kik2, xshift=-0.5cm, yshift=1.25cm] {$+$};
		\draw[-Stealth, thick] (sigma.east) -- (kik2.north);
		\draw[-Stealth, thick] (epsilon.east) -- (kik2.south);
		\draw[-Stealth, thick] (mu.east) -- (kik3.north);
		\draw[-Stealth, thick] (kik2.east) -- (kik3.south);
		\node (zeta) [latentz, right of=kik3] {{$\bZ$}};
		\draw[-Stealth, thick] (kik3.east) -- (zeta.west);
		\node (N) [texting, below of=epsilon, yshift=0.2cm] { $\mathcal{N}(0,1)$};
		\node (q) [texting, below of=zeta, yshift=0.1cm] { $q_{\theta}(\bZ\vert \bX)$};
		\node (decoder) [dec, align=center, right of=zeta, xshift=0.75cm] {Decoder \\ $\tau_{\phi}\left(\bZ \right)$};
		\draw[-Stealth, thick] (zeta.east) -- (decoder.west);
		\node (coeffs2) [coeffrec, right of=decoder, xshift=0.75cm] {{$\bX$}};
		\draw[-Stealth, thick] (decoder.east) -- (coeffs2.west);

	\end{tikzpicture}
    \caption{Architecture of a variational autoencoder (VAE).}
    \label{fig:vae}
\end{figure}
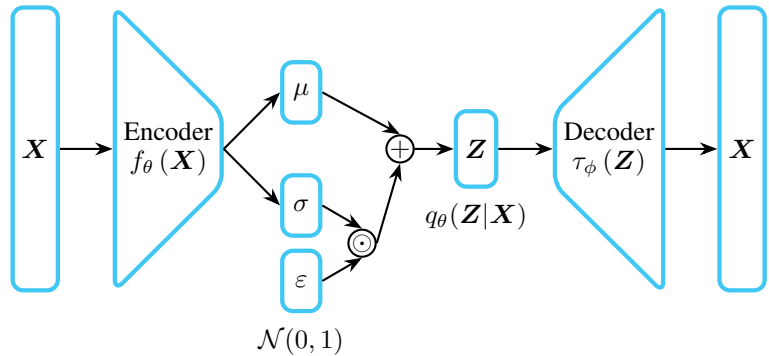

However, sampling cannot be used for the evaluation of the integral in \autoref{eq:phis} due to the computational toll involved and to the integral being analytically intractable.
To address this issue, a \emph{encoder} network is introduced in the \ac{VAE}, also parameterized by $\theta$.
This way, the initially intractable posterior $p(\bZ\vert \bX)$ can be represented approximately via a parameterized distribution $q_{\theta}(\bZ\vert \bX)$, hence creating a mapping from the observation to the latent space.
Next, the variational approximation of the true posterior $p(\bZ)$ results in the following lower bound of the log-likelihood:
\begin{equation}\label{eq:elbo}
    \mathcal{L}(\theta,\phi,\bX)=\E_{q_{\theta}(\bZ\vert \bX)}[log(p_{\phi}(\bX \vert \bZ))]-D_{KL}(q_{\theta}(\bZ\vert \bX)\vert\vert p(\bZ))
\end{equation}
\noindent
where $D_{KL}$ denotes the Kullback-Leibler (KL) divergence metric used to represent distributions' similarity.
The lower bound in \autoref{eq:elbo}, termed the evidence-based lower bound (ELBO), is, in turn, optimized for $\theta$ and $\phi$, the parameters injected in the \emph{encoder} and \emph{decoder} networks. Via maximizing this function, the training aims to i) accurately recover the actual observations from the latent variables via minimizing the reconstruction loss of the \emph{decoder} while ii) minimizing the KL divergence term to increase the similarity between the true and the inferred posterior of the latent space. 
The resulting architecture is visualized in \autoref{fig:vae}.
After properly training the \ac{VAE}, we can sample from the inferred variational distribution $q_{\theta}(\bZ\vert \bX)$ and subsequently utilize the decoder to reproduce the quantities of interest (outputs).

To further optimize \autoref{eq:elbo}, the gradients of ELBO can be estimated using the \emph{re-parameterization trick} \cite{Kingma2014}.
If the approximate posterior is chosen to be a diagonal Gaussian, for example, parameterized by the encoder network, then sampling from it can be re-parameterized as follows;
\begin{align}
\eta &= \mathcal{N}(0,I)\\
q_{\theta}(\bZ\vert \bX) &= \mathcal{N}(\bZ:\mu_{\theta}(\bX),\sigma_{\theta}(\bX))\\
& =\mu_{\theta}(\bX)+\eta\odot\sigma_{\theta}(\bX)
\end{align}
\noindent
in which variable $\eta$ represents a sample drawn from the diagonal Gaussian $\mathcal{N}(0,I)$ and $\mu_{\theta}(\bX),\sigma_{\theta}(\bX)$ denote the mean and standard deviation of the latent space, modeled as outputs of the \emph{encoder} network.
This way, we can recast the approximate posterior as a stochastic draw $\eta$, whereas the deterministic mean and standard deviation values can be estimated by the \emph{encoder}.
This formulation allows us to evaluate the expectation by sampling a standard multivariate Gaussian, while the gradient can be computed in a deterministic manner for each sample, thus enabling back propagation during training.
Further, the KL divergence term can be evaluated analytically due to the choice of a spherical unit Gaussian prior for $p(\bZ)$ \cite{Kingma2014}.
This adaptation gives the following differentiable loss function, expressed per sample, in which the expectation can be evaluated using $N_v$ samples of the latent vector per observation point:
\begin{equation}
    \mathcal{L}(\theta,\phi,x^i)=\frac{1}{2}\sum^J_{j=1}(1+log((\sigma_j^{(i)})^2)-(\mu_j^{(i)})^2-
    (\sigma_j^{(i)})^2)+\frac{1}{L}\sum_{l=1}^{N_v}log(p_{\phi}(X^i,Z^{i,l})
\end{equation}
where $J$ denotes the latent space size and $Z^{i,j} =\mu_{\theta}(X^i)+\eta^l\odot\sigma_{\theta}(X^i) $.
The total number of latent samples $N_v$ required to evaluate the expectation can vary, but even one sample might be enough as in the original formulation \cite{Kingma2014}.

Regarding the networks for the \emph{encoder} and \emph{decoder} portion of the \ac{VAE} deep neural networks (DNNs) are used.
DNNs are widely used as surrogate models that exploit multiple sequential layers of neural networks to approximate complex functions better than shallow networks \cite{BengioLecun}.
A thorough description of DNNs is beyond the scope of this work, and the interested reader is referred to \cite{Goodfellow2014}.

\subsection{VpROM: A conditional Variational Autoencoder (cVAE)-boosted ROM}

The \ac{VAE} presented in the previous section can be exploited as a generator of local projection-subspaces, and thus local ROMs, via sampling the inferred latent space and employing the \emph{decoder} network.
To inject dependencies on the \ac{ROM} level, the latent space of the \ac{VAE} can be \emph{conditioned} on known parameters, thus allowing sampling of local bases on given system properties or excitation traits \cite{simpson2023vprom}. 
However, this requires actual knowledge of the parameter values to sample the latent space and generate a suitable local \ac{ROM} during prediction mode.
Our work goes one step further to derive a \ac{ROM} framework that assumes no knowledge of the actual parameter values after its deployment; rather, it utilizes a conditional \ac{VAE} component as a generative model that can infer the local basis corresponding to features extracted from the monitored response.

\begin{figure}[h]
    \centering
    \tikzstyle{featrec} = [rectangle, rounded corners, 
draw=cyan!60, ultra thick, minimum width=0.5cm, 
minimum height=0.75cm,
text centered]

\tikzstyle{latent} = [rectangle, rounded corners, 
draw=cyan!60, ultra thick, minimum width=0.5cm, 
minimum height=0.8cm,
text centered]

\tikzstyle{texting} = [rectangle, text centered]

\tikzstyle{coeffrec} = [rectangle, rounded corners, 
draw=cyan!60, ultra thick, minimum width=0.5cm, 
minimum height=4.0cm, text centered]

\tikzstyle{kiklos} = [circle,
draw=black, thick,  minimum size = 0.3cm,inner sep=0pt,
text centered]

\tikzstyle{enc} = [trapezium, shape border rotate=270, rounded corners, 
draw=cyan!60, ultra thick, trapezium stretches=false, 
trapezium left angle=45, 
trapezium right angle=45, 
minimum width=1.5cm, 
minimum height=1.5cm, text centered]

\tikzstyle{dec} = [trapezium, shape border rotate=90, rounded corners, 
draw=cyan!60, ultra thick, trapezium stretches=false, 
trapezium left angle=45, 
trapezium right angle=45, 
minimum width=1.5cm, 
minimum height=1.5cm, text centered]

	\begin{tikzpicture}
		\node (coeffs) [coeffrec] {{$\bX$}};
        \node (features) [featrec, above of=coeffs, yshift=1.5cm] {{$\bW$}};
		\node (kik) [kiklos, right of=coeffs, xshift=0.15cm, yshift=0.5cm] {$+$};
		\draw[-Stealth, thick] (features.east) -| (kik.north);
		\draw[-Stealth, thick] (coeffs.east) -| (kik.south);
		\node (encoder) [enc, align=center, right of=kik, xshift=0.5cm] {Encoder \\ $f_{\theta}\left( \bX \right)$};
		\draw[-Stealth, thick] (kik.east) -- (encoder.west);
		\node (mu) [latent, right of=encoder, xshift=0.75cm, yshift=0.75cm] {$\mu$};
		\node (sigma) [latent, below of=mu, yshift=-0.5cm] {$\sigma$};
		\node (epsilon) [latent, below of=sigma] {$\varepsilon$};
		\draw[-Stealth, thick] (encoder.east) -- (mu.west);
		\draw[-Stealth, thick] (encoder.east) -- (sigma.west);
		\node (kik2) [kiklos, right of=sigma, xshift=-0.2cm, yshift=-0.5cm] {$\odot$};
		\node (kik3) [kiklos, right of=kik2, xshift=-0.5cm, yshift=1.25cm] {$+$};
		\draw[-Stealth, thick] (sigma.east) -- (kik2.north);
		\draw[-Stealth, thick] (epsilon.east) -- (kik2.south);
		\draw[-Stealth, thick] (mu.east) -- (kik3.north);
		\draw[-Stealth, thick] (kik2.east) -- (kik3.south);
		\node (zeta) [latent, right of=kik3] {{$\bZ$}};
		\node (cond) [featrec, above of=zeta] {{$\bW$}};
		\draw[-Stealth, thick] (kik3.east) -- (zeta.west);
		\node (N) [texting, below of=epsilon, yshift=0.2cm] { $\mathcal{N}(0,1)$};
		\node (q) [texting, below of=zeta, yshift=0.2cm] { $q_{\theta}\left(\mathbf{Z}|{\mathbf{X} }\right)$};
		\node (kik4) [kiklos, right of=zeta] {$+$};
		\node (decoder) [dec, align=center, right of=kik4, xshift=0.5cm] {Decoder \\ $\tau_{\theta}(\bZ\vert \bX)$};
		\draw[-, thick] (zeta.east) -- (kik4.west);
		\draw[-Stealth, thick] (cond.east) -| (kik4.north);
		\draw[-Stealth, thick] (kik4.east) -- (decoder.west);
		\node (coeffs2) [coeffrec, right of=decoder, xshift=0.5cm] {{$\bX$}};
		\draw[-Stealth, thick] (decoder.east) -- (coeffs2.west);

	\end{tikzpicture}
    \caption{The architecture of a cVAE. The conditioning features $\bW$ are injected via concatenation with the input vector $\bX$ and the latent space $\bZ$. The input refers to the \ac{ROM} basis coefficients $\bX$ in \autoref{eq:coeffs}}
    \label{fig:cvae}
\end{figure}
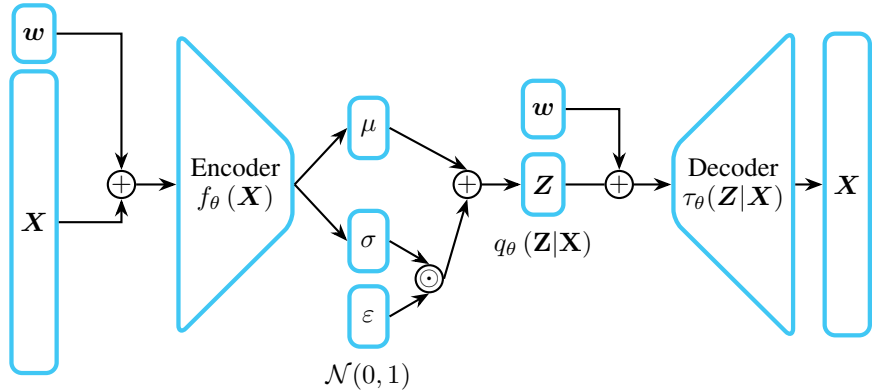

First, we concatenate the conditioning features $\bW$ with the inputs of the \ac{VAE} $\bX$, and the latent space $\bZ$.
The inputs $\bX$ represent the coefficient matrices of the local \ac{ROM}s formulated in \autoref{CpROM} and \autoref{eq:coeffs}.
In terms of mathematical notation, the \emph{encoder} now approximates the distribution $q_{\theta}(\bZ \vert \bX,\bW)$, whereas the \emph{decoder} network the distribution $p_{\phi}(\bX \vert \bZ,\bW)$.
\autoref{fig:cvae} is a visualization of the \ac{cVAE} architecture and the mapping being learned to relate the features $\bW$ extracted from the monitoring response with the local basis' coefficients $\bX$.

Thus, the system's dependencies as expressed in \autoref{eq:FOM} are captured in variable $\bp$, typically reflected in the (properly selected) features $\bW$.
The relation of $\bp$ to the local bases $\bV$ in \autoref{eq:ROM} is captured via learning the mapping between $\bW$, which indirectly expresses $\bp$, and the coefficients $\bX$ from \autoref{eq:coeffs}, which are a reduced representation of $\bV$.

\subsection{VpROM: Generating New Bases}

\begin{figure}[ht]
    \centering
    \tikzstyle{featrec} = [rectangle, rounded corners, 
draw=cyan!60, ultra thick, minimum width=0.5cm, 
minimum height=0.8cm,
text centered]

\tikzstyle{eps} = [rectangle, rounded corners, 
draw=cyan!60, ultra thick, minimum width=0.5cm, 
minimum height=0.8cm,
text centered]

\tikzstyle{latent} = [rectangle, rounded corners, 
draw=cyan!60, ultra thick, minimum width=0.5cm, 
minimum height=1.5cm,
text centered]

\tikzstyle{texting} = [rectangle, text centered]

\tikzstyle{coeffrec} = [rectangle, rounded corners, 
draw=cyan!60, ultra thick, minimum width=0.5cm, 
minimum height=5.25cm,
text centered]

\tikzstyle{basis} = [rectangle, rounded corners, 
draw=cyan!60, ultra thick, minimum width=1.00cm, 
minimum height=7.0cm,
text centered]

\tikzstyle{dec} = [trapezium, shape border rotate=90, rounded corners, 
draw=cyan!60, ultra thick, trapezium stretches=false, 
trapezium left angle=45, 
trapezium right angle=45, 
minimum width=1cm, 
minimum height=2cm, text centered]

	\begin{tikzpicture}
		\node (epsilon) [eps] {$\varepsilon$};
        \node (cond) [featrec, above of=epsilon] {$\bW$};
		\node (zeta) [latent, right of=epsilon] {{$\bZ$}};
		\draw[-Stealth, thick] (epsilon.east) -- (zeta.west);
        \draw[-Stealth, thick] (cond.east) -- (zeta.west);
		\node (N) [texting, below of=epsilon, yshift=0.2cm] { $\mathcal{N}(0,1)$};
		\node (decoder) [dec, align=center, right of=zeta, xshift=0.5cm] {Decoder \\ $\tau_{\theta}(\bZ\vert \bX)$};
		\draw[-, thick] (zeta.east) -- (decoder.west);
		\node (coeffs2) [coeffrec, right of=decoder, xshift=1.0cm] {{$\Tilde{\bX}$}};
		\draw[-Stealth, thick] (decoder.east) -- (coeffs2.west);
        \node (basisV) [basis, right of=coeffs2, xshift=2.5cm] {{$\Tilde{\bV}$}};
        \draw[-Stealth, thick] (coeffs2.east) -- (basisV.west);
        \node (V) [texting, left of=basisV, xshift=-0.825cm, yshift=0.25cm] {$\Tilde{\bV} = \bV_{global} \Tilde{\bX}$};

	\end{tikzpicture}
    \caption{Architecture of the cVAE in basis generation mode: The prior distribution $\epsilon$ is sampled, and the latent vectors are taken after concatenating with the conditioning vector $\bW$. Dimensionality serves only illustrative purposes.}
    \label{fig:VAE_sample}
\end{figure}
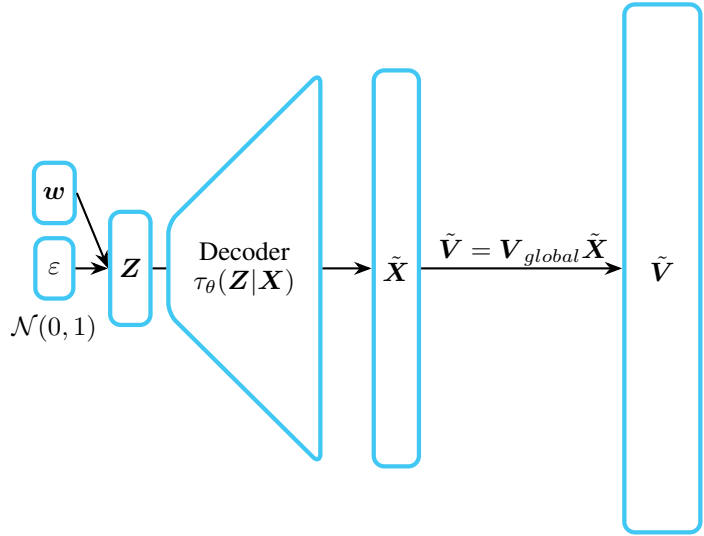

After training the \ac{cVAE}, a generative model is derived that can be sampled to produce the reduced basis coefficients $\bX_i$ in \autoref{eq:coeffs} from the system's monitored response.
Concretely, once the trained \emph{decoder} network is available, predictions of the coefficients can be made via sampling the inferred variational distribution of the latent space and passing it through the \emph{decoder} portion. 
In our case, a diagonal Gaussian distribution is used as shown in \autoref{fig:VAE_sample}. 

Hence, we draw $\epsilon$ from the chosen prior distribution $p(\bZ)$ and concatenate it with the extracted features $\bW$. 
The resulting latent vector is then passed through the \emph{decoder}, resulting in a sample from the observations' distribution $p_{\phi}(\bX \vert \bZ,\bp)$, which is a single realization of the distribution of the predicted coefficients for a given vector of monitoring features.
Next, this process can be repeated to draw multiple such samples and evaluate the mean and standard deviation of the estimated coefficients. 
This generative procedure highlights the significance of minimizing the KL divergence term in the loss function. 
If the KL loss term is low, the resulting approximate posterior $q_{\theta}(\bZ \vert X,p)$ better approaches the prior $p(\bZ)$.

\subsection{Recovery of the actual parameter values}

As visualized in \autoref{fig:VAE_sample}, the \ac{cVAE} component learns the mapping between the monitoring features included in vector $\bW$ and the basis coefficients $\bX$.
In turn, the local basis $\bV$ ban be assembled through \autoref{eq:coeffs}.
To integrate \autoref{eq:ROM}, however, estimating the projection subspace $\bV$ is not enough, as the \ac{ROM} also requires the parameter values $\bp$ that influence the nonlinear restoring force term $\Tilde{\bg}$.
Essentially, the physics-based nature of the derived \ac{ROM} dictates the reconstruction of the forcing term during integration, thus implying the evaluation of the nonlinear effects and the corresponding variables. 
To do so, the parameter values are required.
Thus, the framework necessitates an additional mapping from the vector of monitoring features $\bW$ to the parameter values $\bp$.

To capture the above relationship and further quantify the uncertainty involved, we employ a parametrization approach applied to suitable distributions using neural networks. 
Specifically, let's assume that $\bp \sim \mathcal{N}(\bmu, \bSigma)$.
The multivariate normal distribution $\mathcal{N}$ is used here for the sake of simplicity. 
In addition, we assume that the components of $\bp$, namely the parameters $p_i$, are independent, and thus the covariance matrix of $\mathcal{N}$ is assumed diagonal, and we end up with jointly normally distributed terms. 
Then, we can write:
\begin{equation}
    \bp  \sim \mathcal{N} \left(\bmu,\bSigma \right) \approx \mathcal{\Tilde{N}} \Big( M_{NN}(\bW), \Sigma_{NN}(\bW) \Big)
    \label{eq:param_estim}
\end{equation}
where $M_{NN}, \Sigma_{NN} $ are neural networks that parameterize the mean and standard deviation of the assumed distribution.
The first network termed $M_{NN}$ learns the complex relationship between the parameter vector $\bp$ and the features $\bW$.
This model prediction is treated as the mean value of the joint normal distribution $\mathcal{\Tilde{N}}$, being the most probable outcome.
The second neural network $\Sigma_{NN}$ captures the underlying confidence on the parameter estimation task by quantifying the uncertainty involved. 
Although we have two trainable networks, the goal is to train them together using one single loss function related to the chosen distribution.
Since the probability density function $p \left( \bp | \bmu,\bSigma \right)$ expresses the likelihood when the mean and standard deviation of the respective distribution is given, it can be employed here for the loss function $\mathcal{L}$ of the two neural networks. 
To avoid dealing with the exponential term that appears in the case of the normal distribution, we apply the natural logarithm and then negate the resulting function to recast it as a minimization problem.
Thus, the loss function $\mathcal{L}$ is the negative log-likelihood of observing parameter $\bp$ given $\bW, M_{NN}, \Sigma_{NN}$ as follows:
\begin{equation}
    \mathcal{L} = -\frac{1}{2} \left( \ln\left( \lvert \Sigma_{NN} \rvert \right) + (\bp - M_{NN})^T \Sigma_{NN}^{-1}(\bp - M_{NN}) +  k\ln(2\pi) \right)
    \label{eq:NNloss}
\end{equation}
where $k$ denotes the number of parameters,$\bp \in \subset \bR^k$, and $\Sigma_{NN}=\Sigma_{NN}(\bW)$, $M_{NN}=M_{NN}(\bW)$ with the dependence on the input being dropped for simplicity. 
Overall, the complex parameter estimation task can be carried out with precision due to the increased accuracy and flexibility the neural networks offer. 
At the same time, the parametrization technique on the distribution allows for quantifying the confidence when performing predictions. 

\subsection{Training the VpROM} \label{sec:training}

\begin{figure}[!hbt]
    \centering
    \tikzstyle{featrec} = [rectangle, rounded corners, 
draw=cyan!60, ultra thick, minimum width=0.5cm, 
minimum height=0.75cm,
text centered]

\tikzstyle{latent} = [rectangle, rounded corners, 
draw=cyan!60, ultra thick, minimum width=0.5cm, 
minimum height=0.6cm,
text centered]

\tikzstyle{param} = [rectangle, rounded corners, 
draw=cyan!60, ultra thick, minimum width=0.5cm, 
minimum height=0.8cm,
text centered]

\tikzstyle{texting} = [rectangle, text centered]

\tikzstyle{coeffrec} = [rectangle, rounded corners, 
draw=cyan!60, ultra thick, minimum width=0.5cm, 
minimum height=2.45cm,
text centered]

\tikzstyle{rec} = [rectangle, rounded corners, draw, dashed, minimum width=3.8cm, 
minimum height=2.65cm]

\tikzstyle{rec2} = [rectangle, rounded corners, draw, dashed, minimum width=4.75cm,
minimum height=2.65cm]

\tikzstyle{enc} = [trapezium, shape border rotate=270, rounded corners, 
draw=cyan!60, ultra thick, trapezium stretches=false, 
trapezium left angle=45, 
trapezium right angle=45, 
minimum width=1cm, 
minimum height=1cm, text centered]

\tikzstyle{basis} = [rectangle, rounded corners, 
draw=cyan!60, ultra thick, minimum width=1.00cm, 
minimum height=3.5cm,
text centered]

\tikzstyle{dec} = [trapezium, shape border rotate=90, rounded corners, 
draw=cyan!60, ultra thick, trapezium stretches=false, 
trapezium left angle=45, 
trapezium right angle=45, 
minimum width=1cm, 
minimum height=1cm, text centered]

\tikzstyle{kiklos} = [circle,
draw=black, thick,  minimum size = 0.3cm,inner sep=0pt,
text centered]

\begin{tikzpicture}
	\node (Title) [texting] {\underline{\textit{Training Mode}}};
	\node (Samples) [featrec, below of=Title,  align=center]{Draw parameter \\ samples $\bp$};
	\node (FEM) [texting, below of=Samples, rotate=270,xshift=0.4cm, yshift=0.25cm, align=center]{\textbf{FOM}};
	\node (Snapshots) [featrec, below of=Samples, align=center, yshift = -2.0cm] {Obtain \\ Snapshots};
	\draw[-Stealth, thick] (Samples.south) -- (Snapshots.north);
	\node (Bases)[featrec, right of=Snapshots,  align=center, xshift = 2.75cm]{Local bases $\bV$ \\ Global basis $\bV_{global}$};
	\draw[-Stealth, thick] (Snapshots.east) -- (Bases.west);
	\node(POD) [texting, right of=Snapshots, xshift=0.4cm, yshift=0.2cm ]{\textbf{\small{POD}}};
	\node (coeffs) [coeffrec, right of=Bases, xshift=4.25cm] {{$\bX$}};
	\draw[-Stealth, thick] (Bases.east) -- (coeffs.west);
	\node (lstq) [texting, right of=Bases, align=center, xshift=1.85cm, yshift=0.25cm] {$\bV=\bV_{global}\bX$};
	\node (lstq2) [texting, right of=Bases, align=center, xshift=1.85cm, yshift=-0.25cm] {\textbf{\small{(least-squares)}}};
	\node(Monitor) [featrec, right of=Samples, align=center, xshift=5.0cm]{Monitoring \\ data};
	\draw[-Stealth, thick] (Samples.east) -- (Monitor.west);
	\node (FEM3) [texting, right of=Samples, align=center, xshift=2.15cm, yshift=0.25cm]{\textbf{FOM + Noise}};	
	\node (features) [featrec, above of=coeffs, yshift=0.75cm] {{$\bW$}};
	\draw [-{Latex[length=3mm]}] 
	(Monitor.south) |- node[pos=0.25,name=5] {} (features.west);
	\node(featext)[texting, below of=Monitor,align=center, xshift=1.25cm, yshift=-0.25cm]{\textbf{\small{Feature}} \\ \textbf{\small{extraction}}};
	\node (kik) [kiklos, right of=coeffs, xshift=-0.4cm, yshift=0.5cm] {$+$};
    \draw[-Stealth, thick] (features.east) -| (kik.north);
	\draw[-Stealth, thick] (coeffs.east) -| (kik.south);
    \node (encoder) [enc, align=center, right of=kik, xshift=0.15cm] {Enc.};
    \draw[-Stealth, thick] (kik.east) -- (encoder.west);
	\node (zeta) [latent, right of=encoder, xshift=-0.15cm] {{$\bZ$}};
	\node (featlatent) [featrec, above of=zeta, yshift=-0.25cm] {{$\bW$}};		
	\node (dec) [dec, align=center, right of=zeta, xshift=-0.15cm] {Dec.};	
	\node (coeffs2) [coeffrec, right of=dec, xshift=-0.1cm] {{$\bX$}};
    \draw[-Stealth, thick] (features.east) -| (kik.north);
	\node (param) [param, above of=coeffs2, yshift=1.0cm] {{$\bp$}};
    \draw[-Stealth, thick] (features.north) |- (param.west);
    \node (NN) [texting, above of=featlatent, align=center, yshift=0.5cm, xshift=-0.5cm]{$\mathcal{\Tilde{N}} \Big( M_{NN}(\bW), \Sigma_{NN}(\bW) \Big)$};
    \node (NNN) [texting, above of=featlatent, align=center, yshift=0cm, xshift=-0.5cm]{Eq. \ref{eq:param_estim}};
    \node (rec) [rec, below of=zeta, xshift=0.2cm, yshift=1.0cm] {};
    \node (ref) [texting, below of=rec, align=center, yshift=-0.6cm]{Fig. \ref{fig:cvae}};
    \node (Title2) [texting, below of=Title, yshift=-4.75cm] {\underline{\textit{Prediction Mode}}};
    \node (Inc) [featrec, below of=Title2,  align=center, yshift=-0.65cm]{Incoming \\ monitoring data};
    \node (SamplesT) [featrec, below of=Inc, yshift=-2.2cm, align=center]{Draw testing \\ samples $\Tilde{\bp}$};
    \draw[-Stealth, thick] (SamplesT.north) -- (Inc.south);
    \node (PFEM) [texting, below of=SamplesT, rotate=90, xshift=2.5cm, yshift=0.25cm, align=center]{\textbf{Pert. FOM}};
    \node (PFEM2) [texting, below of=SamplesT, rotate=90, xshift=2.5cm, yshift=-0.25cm, align=center]{\textbf{+ Noise}};
    \node (featuresin) [featrec, right of=Inc, xshift=2.25cm] {{$\Tilde{\bW}$}};
    \draw[-Stealth, thick] (Inc.east) -- (featuresin.west);
    \node(featext2)[texting, right of=Inc, align=center, xshift=1.0cm]{\textbf{\small{Feature}} \\ \textbf{\small{extr.}}};
    \node (zeta2) [latent, below of=featuresin, yshift=0.20cm] {{$\bZ$}};
    \node (dec2) [dec, align=center, right of=zeta2, xshift=-0.15cm] {Dec.};	
    \node (rec2) [rec2, below of=zeta2, xshift=2.0cm, yshift=1.0cm] {};
    \node (ref2) [texting, below of=rec2, align=center, yshift=-0.6cm]{Eq. \ref{fig:VAE_sample}};
    \node (coeffs3) [coeffrec, right of=dec2, xshift=-0.1cm] {{$\Tilde{\bX}$}};
    \node (basisV) [basis, right of=coeffs3, xshift=2.5cm] {{$\Tilde{\bV}$}};
    \draw[-Stealth, thick] (coeffs3.east) -- (basisV.west);
    \node (V) [texting, left of=basisV, xshift=-0.95cm, yshift=0.25cm] {$\Tilde{\bV} = \bV_{global} \Tilde{\bX}$};
    \node (param2) [param, above of=basisV, yshift=1.25cm] {{$\Tilde{\bp}$}};
    \draw[-Stealth, thick] (featuresin.north) |- (param2.west);
    \node (NN2) [texting, left of=param2, align=center, yshift=0.25cm, xshift=-1.5cm]{$\Tilde{\bp} \sim \Tilde{\mathcal{N}}(\Tilde{\bW})$};
    \node (NN3) [texting, left of=param2, align=center, yshift=-0.25cm, xshift=-1.5cm]{Eq. \ref{eq:param_estim}};
    \node (kikB) [kiklos, right of=basisV] {$+$};
    \draw[-, thick] (basisV.east) -- (kikB.west);
    \draw[-Stealth, thick] (param2.east) -| (kikB.north);
    \node (ROMq) [featrec, right of=kikB, xshift=1.25cm, align=center]{$\bq$};
    \draw[-Stealth, thick] (kikB.east) -- (ROMq.west);
    \node (ROM) [texting, right of=kikB, yshift=0.2cm,align=center]{\emph{VpROM}}; 
    \node (ROM2) [texting, right of=kikB, yshift=-0.2cm,align=center]{Eq. \ref{eq:ROM}}; 
    \node (ROMu) [featrec, right of=ROMq, xshift=0.75cm, align=center]{$\bu$}; 
    \draw[-Stealth, thick] (ROMq.east) -- (ROMu.west);
    \node (ROM3) [texting, right of=ROMq,xshift=-0.15cm, yshift=0.2cm,align=center]{$\Tilde{\bV}$};
    \node (ROM3) [texting, right of=ROMq, yshift=-0.2cm,xshift=-0.15cm,align=center]{Eq. \ref{eq:project}};

\end{tikzpicture}
    \caption{Graphical abstract depicting the training and prediction mode of the proposed framework.}
    \label{fig:GraphAbstr}
\end{figure}
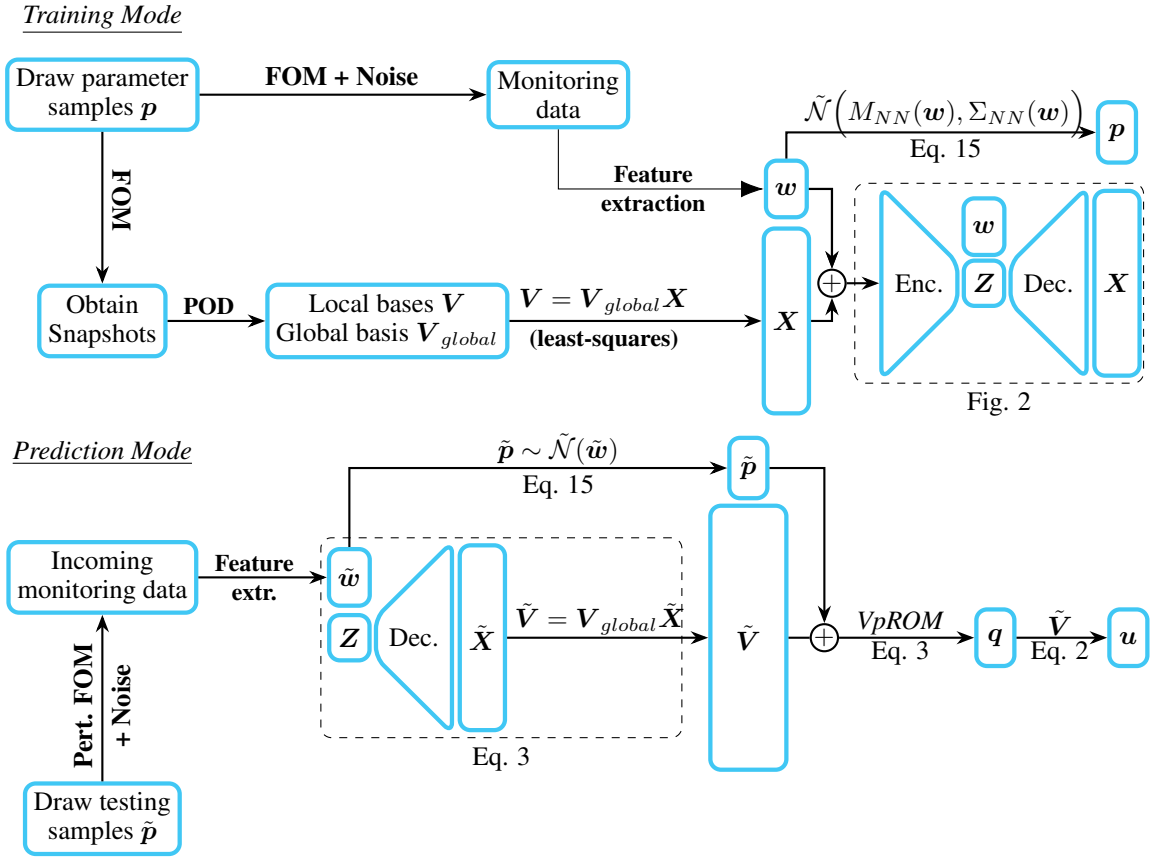

As described in the previous sections, the \emph{VpROM} is trained to generate the coefficient matrices $\bX_i$, which are then used to derive the local \ac{ROM} projection basis based on \autoref{eq:coeffs}.
To achieve this, we require training pairs of monitoring features $\bW_i$ and the corresponding coefficient matrices $\bX_i$.
These training pairs are generated by sampling the parameter vector of the \ac{FOM} and then using each sample as input parameters for a \ac{FOM} snapshot.
Following Equations \ref{eq:Snaps} and \ref{eq:POD}, the generated snapshots are used to assemble the local subspaces $\bV_i$ and the global basis $\bV_{global}$, and hence derive the coefficient matrices following \autoref{eq:coeffs}. 
The vector of monitoring features is derived using a similar procedure but with an independent, perturbed \ac{FOM}.
After evaluating the \emph{perturbed} \ac{FOM} and obtaining the corresponding snapshots for $\bp_i$, measurement noise is added to the acceleration time histories. 
Then, a limited number of nodes is selected as the monitored ones, and feature extraction is implemented in the corresponding acceleration signals to derive $\bW_i$.
The framework's respective training and prediction modes are visualized in \autoref{fig:GraphAbstr}.
Details on the numerical perturbation and measurement setup are provided separately for each validation case study.

Regarding the numerical aspects of the training, our work follows the suggestions in \cite{Simpson2021}. 
However, the loss function of the \ac{cVAE} has been modified to increase robustness and improve generalization. 
Specifically, during the latter stages of the training, the loss function is modified as follows:
\begin{equation}
    \Tilde{\mathcal{L}}(\theta,\phi,x^i)= \mathcal{L}(\theta,\phi,x^i) + \gamma_1* \frac{\lVert \Tilde{\bV}_i\Tilde{\bV}^T_i \hat{\bU} - \hat{\bU} \rVert}{\lVert \hat{\bU} \rVert}+ \gamma_2 * \frac{\lVert \Tilde{\bU}-\hat{\bU} \rVert}{\lVert \hat{\bU} \rVert}
    \label{eq:lossA}
\end{equation}
in which $\Tilde{\bV_i} = \bV_{global} \bX_i$ is the local subspace to be reconstructed, $\hat{\bU}$ represents the displacement (or acceleration) signal of the \ac{FOM}, $\Tilde{\bU}$ is the respective \emph{VpROM} approximation for the system's response and $\gamma_1, \gamma_2$ are regularization parameters to weigh the influence of each additional term.
The second term introduced in \autoref{eq:lossA} represents the reconstruction error 
when using estimated basis $\Tilde{\bV_i}$ to reduce the response signal of the \ac{FOM}.
The third term expresses the ability of the \ac{ROM} to capture and reproduce the underlying dynamics of the \ac{FOM} as it quantifies the error between the \ac{ROM}'s time history estimation $\Tilde{\bu}$, obtained after integrating the response in \autoref{eq:ROM} using the estimated basis $\Tilde{\bV_i}$, and the actual time history $\bu$ from the \ac{FOM}.
Noise can also be added on $\hat{\bU}$ after assembling the local bases $\bV_i$ and before initializing the training process to further refine the training process and improve generalization chances.
Both additional terms in the loss function are considered only at the latter stages as their evaluation increases the computational cost considerably.
Especially for the third term, the \ac{ROM} has to be evaluated, and \autoref{eq:ROM} has to be integrated for every training sample, thus increasing the toll substantially. 

All models and networks were built and trained using Tensorflow and the Adam algorithm \cite{Kingma2015adam}. 
When training the \ac{cVAE} model, the architecture of the network, along with parameters like the number and dimensionality of the layers, the nonlinear activation functions, or the respective rate if dropout exists, need to be defined.
In our work, some of these parameters were also treated as hyper-parameters and were optimized according to common methods such as grid search.
The implementation details, including pre-processing steps, regularization algorithms, hyper-parameter values, and the final architecture of the network, are omitted here for brevity and can be found in \ref{sec:app1}.

\section{Results} \label{sec:Results}

The proposed framework is validated in a series of case studies featuring parametric dependencies in system properties and traits of the excitation. 
We first validate the proposed \ac{cVAE}-boosted \ac{ROM} on an arc structure that experiences geometric nonlinearities due to large deformations.
In this case, the monitoring features are extracted automatically in a naive, uneducated manner from the acceleration time histories using the \emph{tsfresh} python library \cite{christ2018time}.
The \ac{cVAE}-boosted \ac{ROM} is also validated in a large-scale simulation of a kingpin connection featuring computational plasticity, similar to \cite{vlachas2024parametric}. 
Here, the features are extracted following the methodology described in \cite{agathos2022parametric}, where the transmissibility functions are adopted as damage-sensitive features obtained via properly trained ARX variable models \cite{lennart1999system}.
In our case, the measured acceleration signal of each monitored node is approximated by a suitable shallow neural network, utilized as a \ac{NARX} model, which uses as input the respective signal of a neighboring node. 
In turn, the coefficients of all fitted \ac{NARX} models are utilized as the extracted monitoring features that can reflect changes in the system dynamics.  
This \ac{NARX}-based approach is preferred here instead of an uneducated, machine-learning-based extraction due to its inherent ability to treat noise, thus ensuring robustness whilst accurately capturing the dynamics of underlying signals \cite{mai2016surrogate,tatsis2022hierarchical}. 

To obtain the aforementioned features from monitoring measurements experimental data are typically used.
However, due to the limited availability of such information, in our case, simulated measurements are utilized and obtained via an independent numerical \ac{FE} model.
To differentiate this source of monitoring data from the \ac{FOM} used for assembling the \ac{ROM}, thus avoiding a so-called "inverse crime" \cite{kamariotis2023framework}, the following measures are taken:
\begin{itemize}
    \item The \ac{FE} model is assembled in \emph{ABAQUS} \cite{abaqus2023abaqus} instead of the in-house developed Python code that was used to derive the \ac{ROM} with a finer discretization than the one used for the \ac{ROM}s. 
    \item A random perturbation of the Young Modulus $E$ is introduced by drawing its value for each element from a normal distribution with a mean equal to the $E$ value of the sample and a standard deviation equal to 5.
    \item Measurement noise equal to 7\% of the root-mean-square value of each acceleration signal is added.
\end{itemize}

The proposed framework's performance is validated by reproducing the time history response of the respective quantity of interest or capturing its spatial distribution along the discretization for a specific time snapshot of the simulation.
The respective measure utilized is: 
\noindent
\begin{equation}
    \epsilon_q = \dfrac{\sqrt{\sum\limits_{i \in \Tilde{N}_{\text{DOF}}} \sum\limits_{j \in \Tilde{N}_{t}} \left(q_{i}^{j}-\Tilde{q}_{i}^{j}\right)^2}}{\sqrt{\sum\limits_{i \in \Tilde{N}_{\text{DOF}}} \sum\limits_{j \in \Tilde{N}_{t}} \left(q_{i}^{j}\right)^2}} \times 100 \%
    \label{eq:errors}
\end{equation}
\noindent where $\Tilde{N}_{\text{DOF}}$ denotes the selected set of \ac{DOF}s from the spatial discretization, $\Tilde{N}_{t}$ a set of selected simulation time steps, $q_{i}$ is the \ac{FOM} quantity of interest at \ac{DOF} $i$, and $\Tilde{q}_{i}$ is the respective inferred value using the proposed \ac{ROM}.
In the visualizations of the approximation's quality on capturing the time history response, the evaluation is performed on the degree of freedom with the maximum absolute response, whereas the respective confidence bounds are obtained using three standard deviations.
In addition, the average performance refers to the time history approximation that produces the average error measure in \autoref{eq:errors} over all testing parametric samples, whereas the maximum error approximation to the one with the largest error measure. 

With respect to computational resources, all numerical simulations, including the training of the utilized models, are carried out on an in-house \ac{FE} code and tested on a workstation equipped with an \nth{11} Gen Intel(R) Core(TM) i7-1165G7 processor, running at 2.80GHz, and 32GB of memory.
The utilized \ac{FE} code uses \emph{Newmark} integration for time history analysis, whereas both the \ac{FOM} and \ac{ROM} models use the same mesh and time integration increment. 
The speed-ups reported regarding model evaluation do not include the training phase of the \ac{ROM}, termed offline cost, and refer to the average simulation time reduction obtained for every set of model evaluations (validation or testing) in prediction or online mode.

\subsection{Arc structure with large deformations} \label{sec:arc}

As a first example, our work considers a \ac{FE} model of a two-dimensional arc structure experiencing large deformations, and thus geometric nonlinearities.
This example is similar to the curved beam case study in \cite{mahdiabadi2021non}, which experiences a complex dynamic behavior due to the linear coupling between its transverse and in-plane response. 
This system is chosen as a toy problem to demonstrate the ability of the proposed \ac{ROM} to perform accurate predictions even using a naive, automated feature extraction process, namely the \emph{tsfresh} python package \cite{christ2018time}. 
The\emph{tsfresh} python package has been designed for the efficient extraction of a comprehensive set of time series characteristics.
Its feature selection capabilities aid in identifying the most relevant features for downstream machine learning tasks.

\begin{figure}[!htb]
    \begin{subfigure}{0.485\textwidth} 
    \centering
    \includegraphics[scale=0.8]{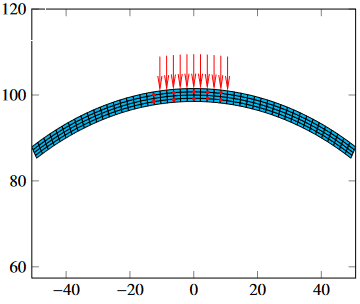}
    \caption{Graphic of the arc structure. \label{fig:arc}}
    \end{subfigure}
    \quad
    \begin{subfigure}{0.485\textwidth}
    \centering
     \includegraphics[scale=0.8]{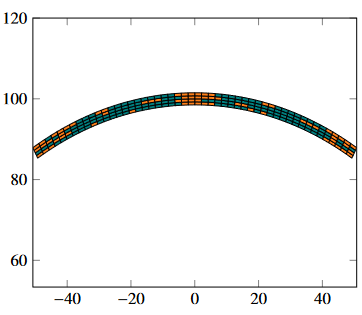}
     \caption{ECSW elements highlighted in orange.\label{fig:hyperarc}}
     \end{subfigure}
\caption{Arc structure: Two-dimensional FE model and example of the ECSW mesh. The monitored nodes are highlighted in red, whereas the ECSW elements are in orange for visualization purposes. \label{fig:arcstruct}}   
\end{figure}

A graphical illustration of the \ac{FE} discretization for the employed arc structure is visualized in \autoref{fig:arc}.
\autoref{fig:hyperarc} demonstrates the subset of elements selected by the ECSW technique discussed in \autoref{sec:hyperreduction} to form the hyper-reduced version of the employed \ac{ROM} framework.
The same subset is used for all samples and simulations. 

\emph{Geometry:} The midline length of the structure is $100mm$ and the respective height difference between the boundaries and the middle cross-section is $15mm$, whereas the height of the cross-section in the middle is $3mm$.
The thickness of the elements is $1mm$ and the width is $10mm$.
For the \ac{FE} discretization nine-noded quadrilaterals are used.

\emph{Excitation:} The structure is excited using a ten second sinusoidal excitation containing 1000 frequency components ranging from $0.1$ to $300$ $Hz$. 
The load is applied in the top surface nodes that span $10m$ from the middle across each side.
These nodes are visualized using red arrows in \autoref{fig:arc}.

\emph{Material properties:} The Poisson ratio is $n=0.28$ for all simulations, whereas the Young modulus $E$ and the density of the material $\rho$ are treated as input parameters following the normal distributions summarized in \autoref{tab:paramsArc}.

\emph{Training}: During training mode, $1000$ parameter samples are drawn from the input space described in \autoref{tab:paramsArc} via \ac{LHS}.
After evaluating the \ac{FOM} and obtaining the training features $\bW$, the \emph{VpROM} can be derived following the methodology described in sections \ref{sec:ReducedOrderModelling} and \ref{sec:VpROM}.
The framework's performance is tested in $250$ samples not included in the training set, where the perturbed, independent \ac{FE} model is evaluated to generate the testing monitoring features $\Tilde{\bW}$. 
In this numerical example, a limited number of ten nodes is assumed to be monitored and the features $\bW, \Tilde{\bW}$ are extracted from the respective acceleration time histories using \emph{tsfresh} \cite{christ2018time}. 
The monitored nodes are highlighted in \autoref{fig:arc} with red dots.
Further details of the training framework can be found in \ref{sec:app1}. 
As a reminder, all the time history response visualizations are evaluated on the degree of freedom with the maximum absolute response, whereas the respective confidence bounds are obtained using three standard deviations.

\begin{table}[!htbp]
\caption{Arc structure: System properties and range of the parameter values of the \ac{ROM}.}
\label{tab:paramsArc}
\centering
\begin{tabular}{l c c c c }
\hline 
Parameter: & Poisson ratio & Young's modulus ($GPa$) & Density $(g/cm^3)$ & Damping \\
Range: & $n=0.28$ & $E \sim \mathcal{N}(210,8^2)$  & $\rho \sim \mathcal{N}(8,0.1)$ & $1\%$ Rayleigh on \nth{1}\& \nth{2} mode \\
\hline
\end{tabular}
\end{table}

\subsubsection*{\textbf{Performance on parameter estimation task}}

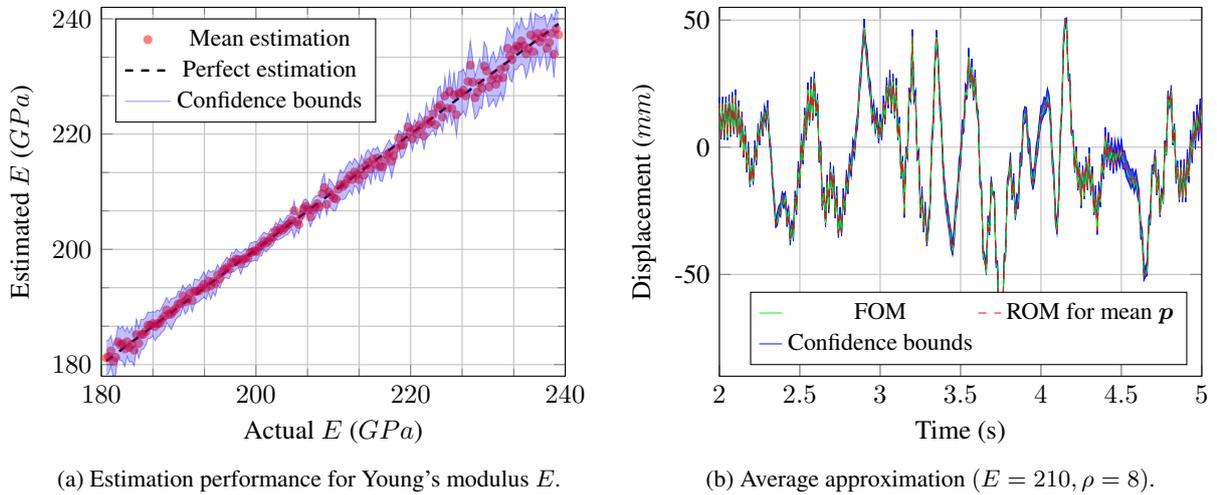
\begin{figure}[!bht]
 \begin{subfigure}[b]{0.49\textwidth}
 \tikzset{mark options={mark size=1.5}}
 \begin{tikzpicture}
		\begin{axis}[width=0.95\textwidth,
		      height=0.80\textwidth,
			xmin = 180,
			xmax = 240,
			ymin = 178,
			ymax = 242,
			xtick = {180, 200, 220, 240},
			ytick = {180, 200, 220, 240},
			xlabel = {Actual $E$ ($GPa$)},
			ylabel = {Estimated $E$ ($GPa$)},
			grid = both,
			minor tick num=2,
			legend style={legend columns=1, legend pos=north west, nodes={font=\fontsize{\figureFontSize pt}{\figureFontSize pt}\selectfont}}
			]
			
			\addplot [draw=red, fill=red, mark=*, only marks, opacity=0.5] table[x=Actual, y=Est] {figures_tikz/ResultsParamEstim_Upd.txt};
			\addplot[name path=F, color = black, dashed, line width=0.95pt] table[x=Actual, y=Actual] {figures_tikz/ResultsParamEstim_Upd.txt};
			\addplot[name path=G, color = blue,opacity=0.4] table[x=Actual, y=Up] {figures_tikz/ResultsParamEstim_Upd.txt};			
			\tikzfillbetween[of=F and G, on layer=ft]{blue, opacity=0.25};
			\addplot[name path=G, color = blue,opacity=0.4] table[x=Actual, y=Down] {figures_tikz/ResultsParamEstim_Upd.txt};			
			\tikzfillbetween[of=F and G, on layer=ft]{blue, opacity=0.25};
			\addlegendentry{Mean estimation}
			\addlegendentry{Perfect estimation}
			\addlegendentry{Confidence bounds}
					
		\end{axis}
		
	\end{tikzpicture}
\caption{Estimation performance for Young's modulus $E$.}
 \label{fig:ParamArc}
 \end{subfigure}
 \begin{subfigure}[b]{0.49\textwidth}
\pgfplotsset{
compat=1.11,
legend image code/.code={
\draw[mark repeat=2,mark phase=2]
plot coordinates {
(0cm,0cm)
(0.15cm,0cm)        
(0.3cm,0cm)         
};%
}
}

\begin{tikzpicture}
	\begin{axis}[
		name = response,
		xmin = 2,
		xmax = 5,
		ymin = -90,
		ymax = 55,
		ytick={-50,0,50},
        yticklabels={-50,0,50},
		xlabel = {Time (s)},
		ylabel = {Displacement ($mm$)},
		grid = both,
		width=0.98\textwidth,
		height=0.80\textwidth,
        legend cell align={center},
		legend style={legend columns=2, legend pos=south east, nodes={font=\fontsize{\figureFontSize pt}{\figureFontSize pt}\selectfont}}
		]
				
        \addplot[name path=F, color = green, line width=0.10pt] table[x=Time, y=FOM] {figures_tikz/Med.txt};
        \addplot[color = red, dashed, line width=0.10pt] table[x=Time, y=ROM]  {figures_tikz/Med.txt};
        \addplot[name path=G, color = blue,opacity=0.90] table[x=Time, y=UpN]  {figures_tikz/Med.txt};	
        \tikzfillbetween[of=F and G,on layer=main]{blue, opacity=0.75};
		\addplot[name path=G, color = blue,opacity=0.90] table[x=Time, y=DownN]  {figures_tikz/Med.txt};		
		\tikzfillbetween[of=F and G,on layer=main]{blue, opacity=0.75};
        \addplot[color = green, line width=0.10pt] table[x=Time, y=FOM] {figures_tikz/Med.txt};
        \addplot[color = red, dashed, line width=0.10pt] table[x=Time, y=ROM]  {figures_tikz/Med.txt};
		\addlegendentry{FOM}
        \addlegendentry{ROM for mean $\bp$}
        \addlegendentry{Confidence bounds}

	\end{axis}
	
\end{tikzpicture}
 \caption{Average approximation $\left(E=210, \rho=8\right)$.}
 \label{fig:AverageDispsArc}
 \end{subfigure}
\caption{Arc structure: Parameter estimation and local \ac{ROM} approximation without the \ac{cVAE} basis generation. The average performance over the samples and the three standard deviation confidence bounds are reported.}
 \label{fig:arc_param_est}
\end{figure}

The framework's performance is reported first with respect to the parameter estimation task, as formulated in \autoref{eq:param_estim}.
The respective prediction obtained via the neural networks $\Tilde{N}$ is visualized in \autoref{fig:ParamArc} for the Young modulus $E$. 
The density $\rho$ estimation is similar. 
The neural networks deliver accurate and robust predictions, as all actual values lie within three standard deviations of the mean, and the confidence bounds are narrow for all testing samples. 
In addition, as expected, $\Tilde{N}$ struggles in the boundaries of the training domain, especially in the upper one as depicted in \autoref{fig:ParamArc} where the confidence bounds grow larger.

The displacement response estimation is visualized in \autoref{fig:AverageDispsArc} for a sample where $\Tilde{N}$ delivers an average quality approximation. 
The respective time history is obtained after evaluating the local \ac{ROM} without utilizing the \ac{cVAE} component for the mean parameter estimation, whereas the confidence bounds are obtained after drawing fifty samples from the respective distribution captured by $\Tilde{N}$ and evaluating the local \ac{ROM} for the corresponding values.

\subsubsection*{\textbf{Performance of the \emph{cVAE} generator alone without the parameter inference task}}

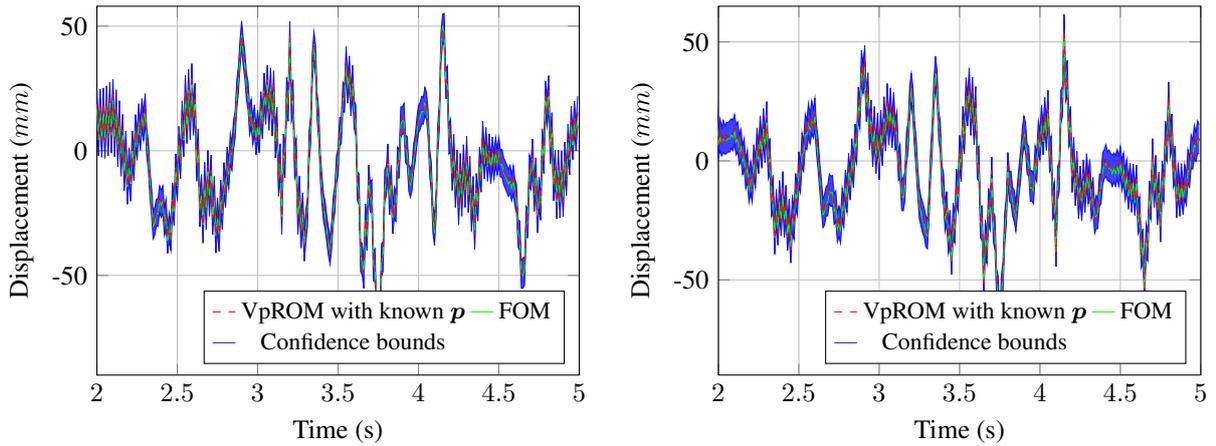
\begin{figure}[!hbt]
    \begin{subfigure}[c]{0.49\textwidth}
        \pgfplotsset{
compat=1.11,
legend image code/.code={
\draw[mark repeat=2,mark phase=2]
plot coordinates {
(0cm,0cm)
(0.15cm,0cm)        
(0.3cm,0cm)         
};%
}
}

\begin{tikzpicture}
	\begin{axis}[
		name = response,
		xmin = 2,
		xmax = 5,
		ymin = -90,
		ymax = 58,
		ytick={-50,0,50},
        yticklabels={-50,0,50},
		xlabel = {Time (s)},
		ylabel = {Displacement ($mm$)},
		grid = both,
		width=0.98\textwidth,
		height=0.80\textwidth,
        legend cell align={center},
		legend style={legend columns=2, legend pos=south east, nodes={font=\fontsize{\figureFontSize pt}{\figureFontSize pt}\selectfont}}
		]

        \addplot[color = red, dashed, line width=0.10pt] table[x=Time, y=ROM]  {figures_tikz/BasisUQ_Med.txt};				
        \addplot[name path=F, color = green, line width=0.10pt] table[x=Time, y=FOM] {figures_tikz/BasisUQ_Med.txt};
        \addplot[name path=G, color = blue,opacity=0.90] table[x=Time, y=Up]  {figures_tikz/BasisUQ_Med.txt};	
        \tikzfillbetween[of=F and G,on layer=main]{blue, opacity=0.75};
		\addplot[name path=G, color = blue,opacity=0.90] table[x=Time, y=Down]  {figures_tikz/BasisUQ_Med.txt};		
		\tikzfillbetween[of=F and G,on layer=main]{blue, opacity=0.75};
        \addplot[color = green, line width=0.10pt] table[x=Time, y=FOM] {figures_tikz/BasisUQ_Med.txt};
        \addplot[color = red, dashed, line width=0.10pt] table[x=Time, y=ROM]  {figures_tikz/BasisUQ_Med.txt};
        \addlegendentry{VpROM with known $\bp$ }
        \addlegendentry{FOM}
        \addlegendentry{Confidence bounds}

	\end{axis}
	
\end{tikzpicture}
        \caption{Average approximation $\left(E=210, \rho=8\right)$.}  \label{fig:average_basis}
    \end{subfigure}     
    \begin{subfigure}[c]{0.49\textwidth}
        \pgfplotsset{
compat=1.11,
legend image code/.code={
\draw[mark repeat=2,mark phase=2]
plot coordinates {
(0cm,0cm)
(0.15cm,0cm)        
(0.3cm,0cm)         
};%
}
}

\begin{tikzpicture}
	\begin{axis}[
		name = response,
		xmin = 2,
		xmax = 5,
		ymin = -90,
		ymax = 65,
		ytick={-50,0,50},
        yticklabels={-50,0,50},
		xlabel = {Time (s)},
		ylabel = {Displacement ($mm$)},
		grid = both,
		width=0.98\textwidth,
		height=0.80\textwidth,
        legend cell align={center},
		legend style={legend columns=2, legend pos=south east, nodes={font=\fontsize{\figureFontSize pt}{\figureFontSize pt}\selectfont}}
		]

        \addplot[color = red, dashed, line width=0.10pt] table[x=Time, y=ROM]  {figures_tikz/BasisUQ_Max.txt};				
        \addplot[name path=F, color = green, line width=0.10pt] table[x=Time, y=FOM] {figures_tikz/BasisUQ_Max.txt};
        \addplot[name path=G, color = blue,opacity=0.90] table[x=Time, y=Up]  {figures_tikz/BasisUQ_Max.txt};	
        \tikzfillbetween[of=F and G,on layer=main]{blue, opacity=0.75};
		\addplot[name path=G, color = blue,opacity=0.90] table[x=Time, y=Down]  {figures_tikz/BasisUQ_Max.txt};		
		\tikzfillbetween[of=F and G,on layer=main]{blue, opacity=0.75};
        \addplot[color = green, line width=0.10pt] table[x=Time, y=FOM] {figures_tikz/BasisUQ_Max.txt};
        \addplot[color = red, dashed, line width=0.10pt] table[x=Time, y=ROM]  {figures_tikz/BasisUQ_Max.txt};
		\addlegendentry{VpROM with known $\bp$ }
        \addlegendentry{FOM}
        \addlegendentry{Confidence bounds}

	\end{axis}
	
\end{tikzpicture}
        \caption{Maximum error approximation $\left(E=231, \rho=7.79\right)$.}\label{fig:worst_basis}
    \end{subfigure}
	\caption{Arc structure: Response approximation utilizing the VpROM when assuming knowledge of the parameters $\bp$. The average and maximum error performance over the samples is reported with three standard deviation bounds.}
	\label{fig:arc_basis_est}
\end{figure}

The performance of the \emph{cVAE} in estimating the local basis $\bV$ and the displacement response via integrating the derived local \ac{ROM} is visualized in \autoref{fig:arc_basis_est}.
The parameter values are considered known and the respective inference task is not incorporated here. 
To obtain the confidence bounds illustrated in \autoref{fig:arc_basis_est}, we rely on the latent space distribution of the \ac{cVAE} component of the \emph{VpROM}.
Specifically, for given features $\Tilde{\bW}$, one hundred samples are drawn from the inferred distribution of the latent space and then passed from the decoder portion of the network.
This results in one hundred projection bases and allows the \emph{VpROM} to propagate the dynamics for each one in parallel, thus obtaining the corresponding bounds.
The actual response prediction of the \emph{VpROM} is obtained using the latent space's corresponding mean value. 

The uncertainty in \autoref{fig:arc_basis_est} is higher compared to the the uncertainty due to the parameter inference task in \autoref{fig:arc_param_est}, indicating that the local basis estimation is a more complex task in this case. 
Nonetheless, the actual \ac{FOM} response lies within the confidence bounds provided by the \ac{cVAE} decoder both for the average approximation in \autoref{fig:average_basis} and the maximum error approximation in \autoref{fig:worst_basis}, demonstrating the robustness of the proposed \emph{VpROM}.

Next, the parameter inference task is incorporated into the \ac{cVAE} basis generation to derive the hyper-reduced \emph{VpROM} framework. 

\subsubsection*{\textbf{Performance of the fully-assembled hyper-reduced \emph{VpROM} framework}}

\begin{figure}[!bht]
\begin{tikzpicture}[spy using outlines=
	{rectangle, magnification=2.5, anchor = center, connect spies}]
	\begin{axis}[
		name = response,
		xmin = 2,
		xmax = 5,
		ymin = -82,
		ymax = 60,
		xlabel = {Time (s)},
		ylabel = {Displacement ($mm$)},
		grid = both,
		width=0.75\textwidth,
		height=0.35\textwidth,
		legend style={legend columns=-1, legend pos=south east}
		]
		
		\addplot[name path=F, color = green, line width=0.10pt] table[x=Time, y=FOM] {figures_tikz/AllUQ_Med.txt};
		\addplot[color = red, dashed, line width=0.10pt] table[x=Time, y=ROM]  {figures_tikz/AllUQ_Med.txt};
		\addplot[name path=G, color = blue,opacity=0.90] table[x=Time, y=Up]  {figures_tikz/AllUQ_Med.txt};			
		\tikzfillbetween[of=F and G,on layer=main]{blue, opacity=0.75};
		\addplot[name path=G, color = blue,opacity=0.90] table[x=Time, y=Down]  {figures_tikz/AllUQ_Med.txt};		
		\tikzfillbetween[of=F and G,on layer=main]{blue, opacity=0.75};
        \addplot[color = green, line width=0.20pt] table[x=Time, y=FOM] {figures_tikz/AllUQ_Med.txt};
		\addplot[color = red, line width=0.10pt] table[x=Time, y=ROM]  {figures_tikz/AllUQ_Med.txt};
		\addlegendentry{FOM}
		\addlegendentry{(Hyper-reduced) VpROM}
        \addlegendentry{Confidence bounds}
		
		\coordinate (spypoint) at (axis cs:3.00,22);
		
		\end{axis}
		
		\path node[anchor= center] (magnifyglass) at (0.8\textwidth,0.13\textwidth) {};
		
		\spy [black, every spy on node/.append style={ultra thick}, width=0.25\textwidth, height=0.25\textwidth] on (spypoint)
		in node[fill=white] at (magnifyglass);

\end{tikzpicture}
\captionsetup{skip=-5pt}
\caption{Arc structure: Average quality of the \emph{VpROM} response approximation. The average performance over the parametric samples is reported $\left(E=210, \rho=8\right)$ with three standard deviation confidence bounds.}
\label{fig:Arc_UQ_THsmed}
\end{figure}
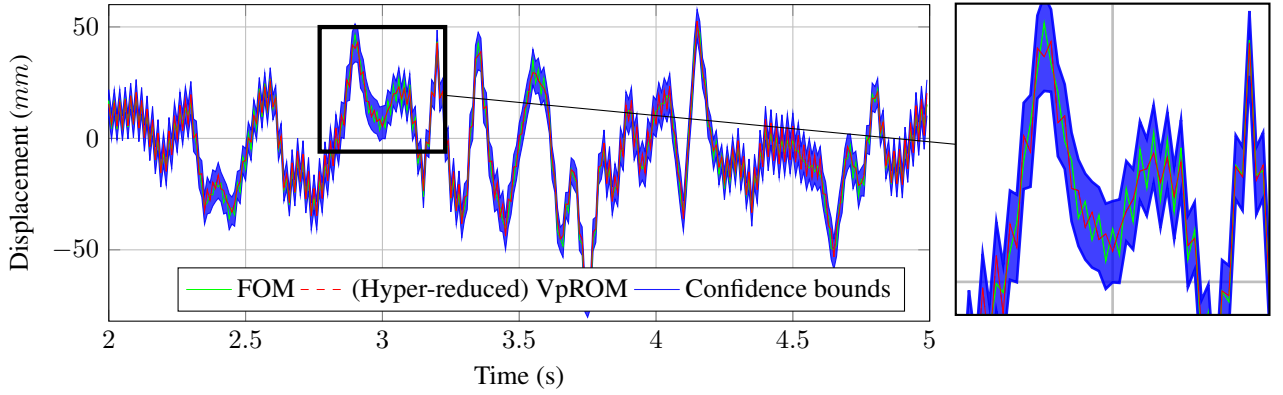

The fully-assembled \emph{VpROM} framework as proposed in \autoref{sec:VpROM} is assembled, utilizing both the parameter inference neural networks and the \ac{cVAE} basis generation component. 
The resulting low-order model is additionally equipped with hyper-reduction and is evaluated in the testing parameter samples. 
The average displacement response approximation is illustrated in \autoref{fig:Arc_UQ_THsmed}

\begin{figure}[!htb]
\begin{tikzpicture}[spy using outlines=
	{rectangle, magnification=2.5, anchor = center, connect spies}]
	\begin{axis}[
		name = response,
		xmin = 2,
		xmax = 5,
		ymin = -82,
		ymax = 60,
		xlabel = {Time (s)},
		ylabel = {Displacement ($mm$)},
		grid = both,
		width=0.75\textwidth,
		height=0.4\textwidth,
		legend style={legend columns=-1, legend pos=south east}
		]
		
		\addplot[name path=F, color = green, line width=0.10pt] table[x=Time, y=FOM] {figures_tikz/AllUQ_Max.txt};
		\addplot[color = red, dashed, line width=0.10pt] table[x=Time, y=ROM]  {figures_tikz/AllUQ_Max.txt};
		\addplot[name path=G, color = blue,opacity=0.90] table[x=Time, y=Up]  {figures_tikz/AllUQ_Max.txt};			
		\tikzfillbetween[of=F and G,on layer=main]{blue, opacity=0.75};
		\addplot[name path=G, color = blue,opacity=0.90] table[x=Time, y=Down]  {figures_tikz/AllUQ_Max.txt};		
		\tikzfillbetween[of=F and G,on layer=main]{blue, opacity=0.75};
        \addplot[color = green, line width=0.20pt] table[x=Time, y=FOM] {figures_tikz/AllUQ_Max.txt};
		\addplot[color = red, line width=0.10pt] table[x=Time, y=ROM]  {figures_tikz/AllUQ_Max.txt};
		\addlegendentry{FOM}
		\addlegendentry{(Hyper-reduced) VpROM}
        \addlegendentry{Confidence bounds}
		
		\coordinate (spypoint) at (axis cs:4.75,5);
		
		\end{axis}
		
		\path node[anchor= center] (magnifyglass) at (0.8\textwidth,0.15\textwidth) {};
		
		\spy [black, every spy on node/.append style={ultra thick}, width=0.25\textwidth, height=0.30\textwidth] on (spypoint)
		in node[fill=white] at (magnifyglass);

\end{tikzpicture}
\captionsetup{skip=-5pt}
\caption{Arc structure: Quality of the \emph{VpROM} approximation. The maximum error approximation over the parametric samples is reported $\left(E=231, \rho=7.79\right)$ with three standard deviation confidence bounds.}
	\label{fig:Arc_UQ_THsmax}
\end{figure}
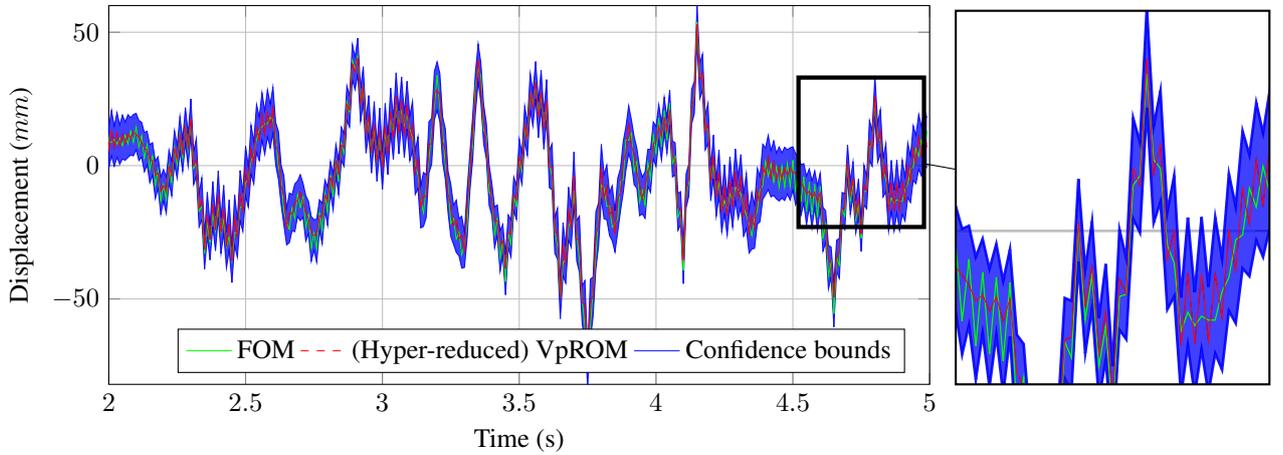

The accuracy of the proposed \emph{VpROM} is considerably lower compared to previous work utilizing the \ac{cVAE} framework \cite{simpson2023vprom} or similar approaches in machine-learning assisted model order reduction \cite{barnett2023neural, fresca2022pod, azzi2023acceleration}.
Specifically, due to the discrepancy when estimating the physical parameters, here representing the material properties and the basis generation, the obtained approximation is not identical to the \ac{FOM} response.
This is visualized in \autoref{fig:Arc_UQ_THsmed}, where the average performance of the \emph{VpROM} over the parametric samples is reported.
However, the respective confidence bounds are narrow, thus implying a \ac{ROM} able to provide accurate response envelopes with high confidence.
The same applies to the maximum error approximation, visualized in \autoref{fig:Arc_UQ_THsmax}. 
As expected, this sample is located in the boundaries of the training domain, where the framework struggles more.
Thus, the uncertainty bounds in \autoref{fig:Arc_UQ_THsmax} are larger, indicating a higher uncertainty than the average quality approximation in \autoref{fig:Arc_UQ_THsmax}.

The aforementioned inability to provide very high precision estimates is addressed here without exhaustively sampling the parameter domain, increasing the number of monitoring sensors, or increasing the computational resources required during training in an intractable manner.
Instead, the proposed framework provides confidence bounds on the delivered estimation, thus quantifying the uncertainty involved and offering accurate response envelopes. 
The prediction confidence remains high even in the worst-case scenario in \autoref{fig:Arc_UQ_THsmax} as a) the actual response lies within the estimated bounds at all times and b) the discrepancies between the actual response and the upper and lower time history estimation are small compared to the response magnitude.

\begin{table}[!htb]
    \centering
	\caption{Arc structure: Performance summary of the \emph{VpROM} and its hyper-reduced version. Accuracy is reported with respect to displacements ($\epsilon_u$). Timings exclude the training phase and refer to online model evaluations.}
	\label{tab:ErrorsArc}
		\begin{tabular} { |l|  c|  c| c|}
        \hline
         & \ac{FOM} & \emph{VpROM} & Hyper-reduced \emph{VpROM}\\
         \hline
         model size             &    510       &      16     &      16     \\
         elements                &    200        &    200  &      68       \\ 
         Maximum error $\epsilon_u$ (\%)        &    0.00       &    13.62  &      18.60     \\
         Mean error $\epsilon_u$ (\%)           &    0.00       &    10.28  &      16.54     \\
         Training time (s)       &    -       &    31456   &      31456    \\
         Solution time (s)       &    1461       &    1072   &      100    \\
         Speedup ($t_{FOM}/t_{ROM}$)                  &    1.00       &    1.36   &      14.56   \\
	    \hline
    \end{tabular}
\end{table}

Thus, the employed \emph{VpROM} framework faces several sources of error and uncertainty coming from model parameters and basis estimation from noisy monitoring features, basis truncation, and the hyper-reduction approximation.
Nonetheless, it derives a robust and sufficiently accurate response estimation.
At the same time, the framework leads to a considerable model evaluation speedup, along with substantial computational toll reduction, as reported in \autoref{tab:ErrorsArc}.
This implies that the proposed approach is potentially highly useful for digital twinning and \ac{SHM} applications. 
The framework is validated next in a large-scale example featuring plasticity.

\subsection{Kingpin system with plasticity} \label{sec:kingpin}

The performance of the proposed \emph{VpROM} framework is assessed next for the case of a large-scale kingpin connection system, where the weld region and the kingpin component are characterized by plasticity.

\begin{figure}[!htb]
	\begin{subfigure}[b]{0.5\textwidth}
		\centering
		\includegraphics[scale=0.20]{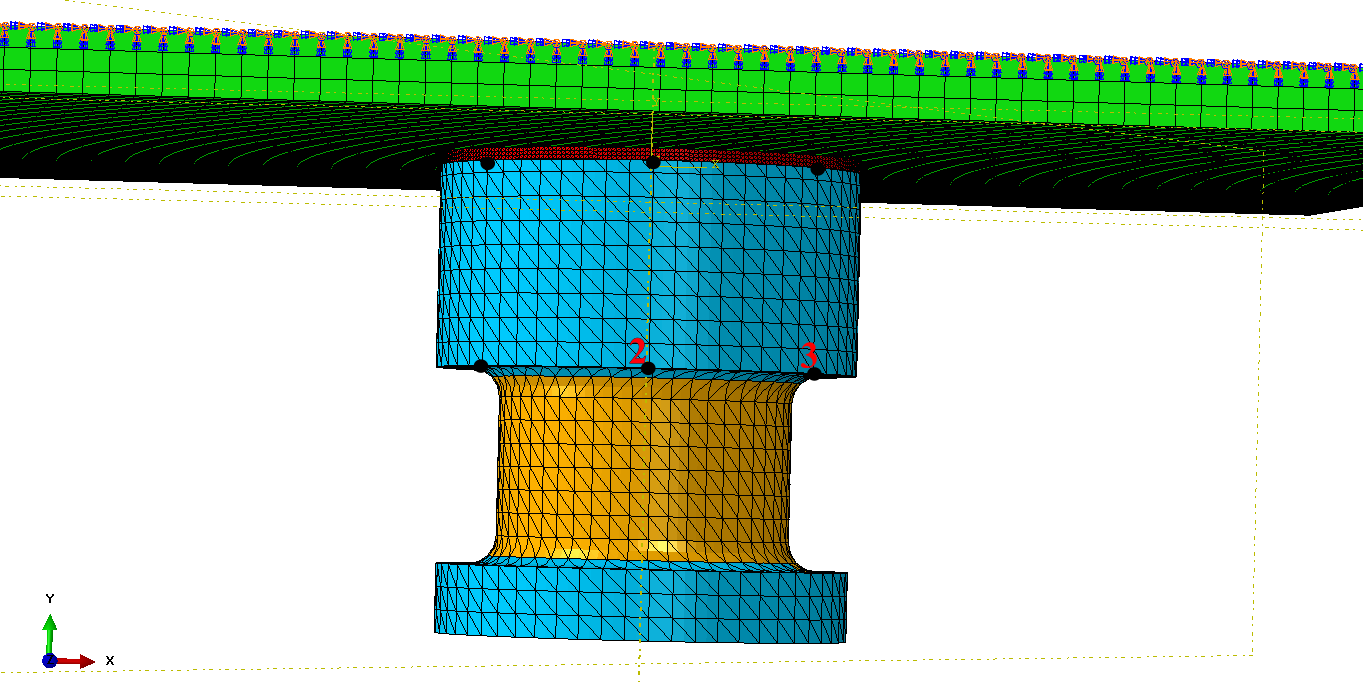}
		\caption{\ac{FE} model for the kingpin connection.}
		\label{fig:geometryKP}
	\end{subfigure}
    \begin{subfigure}[b]{0.5\textwidth}
		\centering
		\includegraphics[scale=0.25]{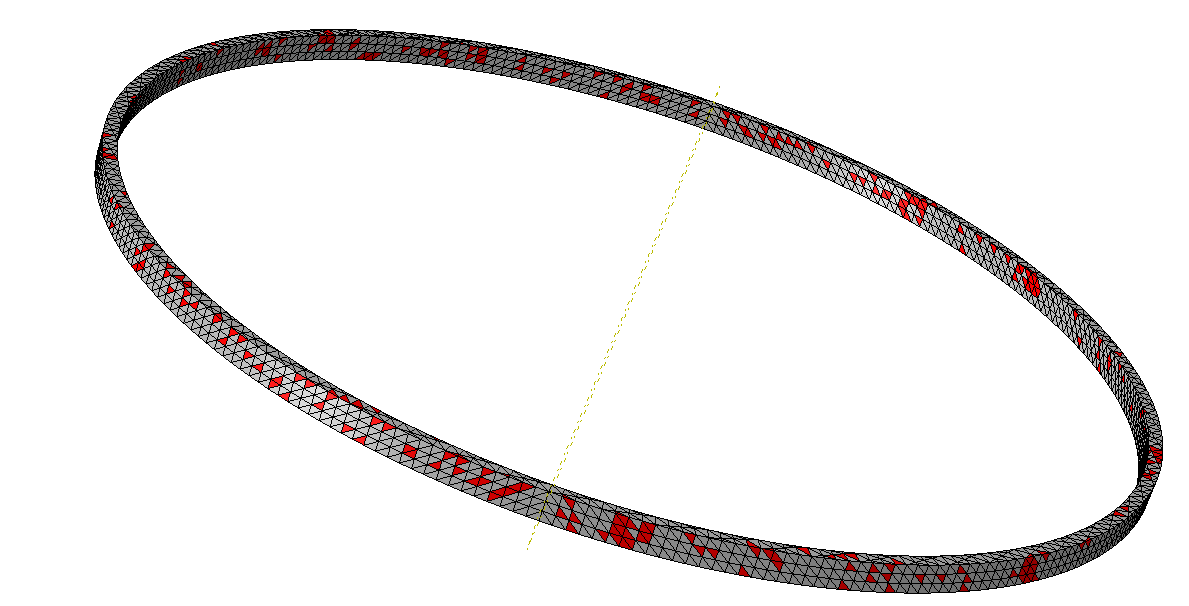}
		\caption{ECSW mesh for the welding region highlighted in red.}
		\label{fig:hyperF}
	\end{subfigure}
	\caption{Kingpin system: The \ac{FE} model for the kingpin connection and the ECSW mesh for the welding component. The excitation domain is depicted in orange in the \ac{FE} model and the monitored nodes are in black.}   
    \label{fig:geometry}
\end{figure}

\emph{Geometry:} A graphic of the \ac{FE} model used for this case study is visualized in \autoref{fig:geometryKP}. 
The kingpin is cyan-colored and is connected to a steel plate illustrated with green. 
The respective welding region materializing the connection is highlighted in red. 
The \ac{FE} model is discretized using $12150$ tetrahedral elements for the green upper plate of the system, whereas its top surface is considered fully bounded, as depicted in \autoref{fig:geometryKP}.
The welding region consists of $11543$ quadratic tetrahedral elements, and the cyan-colored kingpin has $28351$ elements, respectively.

\emph{Material Properties:} Linear elastic material properties are assumed for the upper plate, while the kingpin and the circular welding region follow a von Mises plasticity rule.
In addition, a lower yielding limit is modeled for the welding component, assuming it represents a weak spot in the system.
The respective geometric and material properties of the system that are not treated as parameters are summarized in \autoref{tab:KingpinSetUp}, whereas the yielding stresses of the kingpin and the welding component follow the respective distributions in \autoref{tab:KingpinParams}.
Rayleigh proportional damping is also assumed, corresponding to 1\% modal damping, ascribed to the first and second global modes of the system.

\begin{table}[!htb]
\caption{Kingpin system: Geometry and material properties not treated as varying parameters.}
\label{tab:KingpinSetUp}
\centering
\begin{tabular}{l c c c c}
\hline 
& Young Modulus & Density & Poisson ration  & Dimensions\\
Upper Plate: & $E=200$ $GPa$ & $\rho=8100$ $kg/m^3$ & $\nu=0.30$ & Thickness=$10mm$, Edge=$300mm$ \\
Welding region: & $E=198$ $GPa$ & $\rho=7850$ $kg/m^3$ & $\nu=0.29$ &  Thickness=$2mm$, Radius=$74mm$ \\
Kingpin: & $E=210$ $GPa$ & $\rho=7850kg/m^3$ & $\nu=0.32$ &  Height=$83mm$ \\
\hline
\end{tabular}
\end{table}

\emph{Excitation:} The system is excited at the yellow region of the kingpin depicted in \autoref{fig:geometryKP}.
This domain represents the region where the connection of a vehicle's (truck) trailer to its tractor materializes; this drives the trailer forward through the kingpin and the corresponding fifth-wheel connection. 
Thus, the modeled induced excitation represents the traction and pulling forces or vibrations originating from the trailer during driving or steering.
For this reason, we consider a parameterized acceleration signal as input excitation, applied to the discretization nodes of the highlighted orange region in \autoref{fig:geometryKP}.
Specifically, the imposed input acceleration is modeled as a stochastic multi-sinusoidal signal.
Its amplitude $A$ is assumed to follow a normal distribution, and its direction in the horizontal (x-y) plane of motion (no component along the height of the system) is modeled through angle $\theta$ that also follows a normal distribution.
The respective distributions are summarized in \autoref{tab:KingpinParams}.

\begin{table}[!htb]
\caption{Kingpin system: Parametric dependencies of the \ac{FOM} model of the connection.}
\label{tab:KingpinParams}
\centering
\begin{tabular}{l c c c c c}
\hline 
Parameter: & Excitation amplitude & Excitation direction & Yield stress of weld & Yield stress of kingpin \\
Distribution: & $A \sim \mathcal{N}\left(100,36\right)$ & $\theta \sim \mathcal{N}\left(\frac{\pi}{4},\left(\frac{\pi}{18}\right)^2\right)$ & $f_y \sim \mathcal{N}\left(370,25\right)$ MPa & $f_y \sim \mathcal{N}\left(435,49\right)$ MPa \\
\hline
\end{tabular}
\end{table}

\emph{Training:} Following a similar strategy as for the arc structure, 1000 parameter samples are drawn using the distributions in \autoref{tab:KingpinParams} and \ac{LHS}, and two independent models are evaluated, namely the \ac{FOM} to obtain the snapshots and derive the ROM following \autoref{sec:ReducedOrderModelling}, and the perturbed, independent \ac{FE} model to generate the incoming testing monitoring information.
Validation is performed in a testing set of $200$ samples. 
A limited number of twelve nodes are monitored. 
Half of them are depicted with black dots in \autoref{fig:geometryKP}, whereas the rest six are distributed equivalently on the back side of the \ac{FE} model.

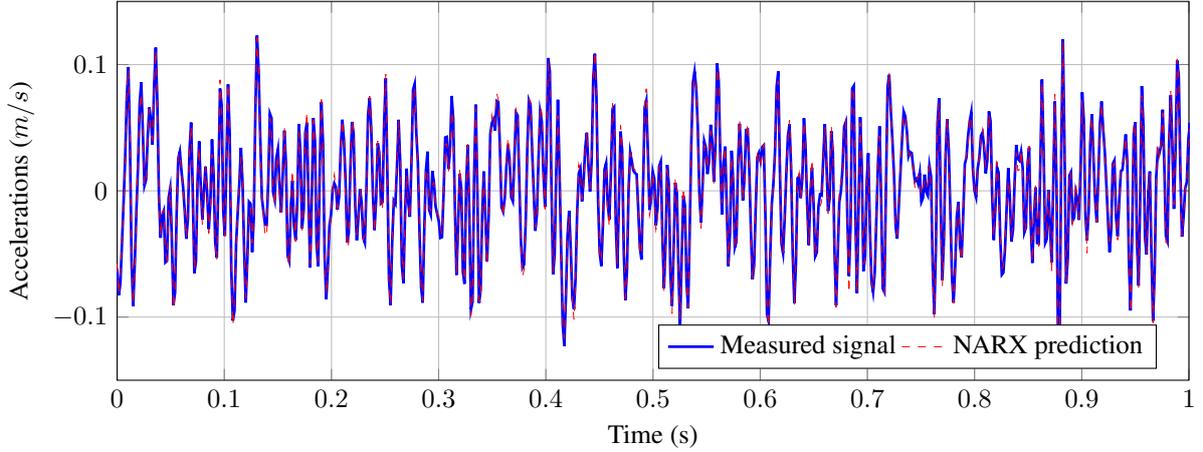
\begin{figure}[!htb]
\centering
\begin{tikzpicture}
    \begin{axis}[
		name = response,
		xmin = 0,
		xmax = 1,
		ymin = -0.15,
		ymax = 0.15,
		xlabel = {Time (s)},
		ylabel = {Accelerations ($m/s$)},
		grid = both,
		width=0.95\textwidth,
		height=0.4\textwidth,
		legend style={legend columns=-1, legend pos=south east}
		]
		
		\addplot[color = blue, line width=1.0pt] table[x=Time, y=FOMn]{./ARX_test.txt};
		\addplot[color = red, dashed, line width=0.25pt] table[x=Time, y=NARX]  {./ARX_test.txt};
		\addlegendentry{Measured signal}
		\addlegendentry{NARX prediction}
		
		\end{axis}
	
\end{tikzpicture}
\caption{Kingpin system: The NARX model prediction of the horizontal acceleration at sensor 3 using the acceleration at sensor 2 as input.}
\label{fig:arxfit}
\end{figure}

The features $\bW, \Tilde{\bW}$ are extracted after fitting a \ac{NARX} model in the acceleration signal of each node pair-wisely, using the incoming signal of the closest neighboring node as input. 
An example of the \ac{NARX} model prediction is visualized in \autoref{fig:arxfit}.
All other \ac{NARX} models employed exhibit a similar performance.
The horizontal acceleration of sensor three is predicted using the monitoring signal of sensor two as input.
The location for both sensors is highlighted in \autoref{fig:geometryKP} with red. 
The \ac{NARX} estimation captures the trend and dynamics of the original signal while exhibiting robustness to measurement noise, thus implying an effective approach for feature extraction.
A time window of three seconds is used, and since each \ac{NARX} model describes the vibration response of the same quantity and not an actual input-output mapping, the time delay parameter for all models is set to zero, and the order of both the AR and X component to $64$.
The coefficients of the \ac{NARX} models are then stacked, and an additional \ac{PCA} decomposition is performed for dimensionality reduction to obtain the features $\bW$.
Further details of the training, the employed models, and network architectures can be found in \ref{sec:app1}.

\subsubsection*{\textbf{Performance on parameter estimation and basis generation in separate}}

\begin{figure}[!htb]
    \begin{subfigure}[c]{0.49\textwidth}
        \tikzset{mark options={mark size=1.5}}
 \begin{tikzpicture}
		\begin{axis}[width=0.95\textwidth,
		      height=0.80\textwidth,
			xmin = 0,
			xmax = 1,
			ymin = -0.05,
			ymax = 1.05,
			xlabel = {Actual normalized $f_y$},
			ylabel = {Estimated normalized $f_y$},
			grid = both,
			minor tick num=2,
			legend style={legend columns=1, legend pos=north west, nodes={font=\fontsize{\figureFontSize pt}{\figureFontSize pt}\selectfont}}
			]
			
			\addplot [draw=red, fill=red, mark=*, only marks, opacity=0.5] table[x=Actual, y=Est] {figures_tikz/ParamYield.txt};
			\addplot[name path=F, color = black, dashed, line width=0.95pt] table[x=Actual, y=Actual] {figures_tikz/ParamYield.txt};
			\addplot[name path=G, color = blue,opacity=0.4] table[x=Actual, y=Up] {figures_tikz/ParamYield.txt};			
			\tikzfillbetween[of=F and G, on layer=ft]{blue, opacity=0.25};
			\addplot[name path=G, color = blue,opacity=0.4] table[x=Actual, y=Down] {figures_tikz/ParamYield.txt};			
			\tikzfillbetween[of=F and G, on layer=ft]{blue, opacity=0.25};
			\addlegendentry{Mean estimation}
			\addlegendentry{Perfect estimation}
			\addlegendentry{Confidence bounds}
					
		\end{axis}
		
	\end{tikzpicture}
        \caption{Parameter estimation for weld's yield stress$f_y$.}  \label{fig:kingpin_param_yield}
    \end{subfigure}     
    \begin{subfigure}[c]{0.49\textwidth}
        \tikzset{mark options={mark size=1.5}}
 \begin{tikzpicture}
		\begin{axis}[width=0.95\textwidth,
		      height=0.80\textwidth,
			xmin = 0,
			xmax = 1,
			ymin = -0.2,
			ymax = 1.20,
			xlabel = {Actual normalized $\theta$},
			ylabel = {Estimated normalized $\theta$},
			grid = both,
			minor tick num=2,
			legend style={legend columns=1, legend pos=north west, nodes={font=\fontsize{\figureFontSize pt}{\figureFontSize pt}\selectfont}}
			]
			
			\addplot [draw=red, fill=red, mark=*, only marks, opacity=0.5] table[x=Actual, y=Est] {figures_tikz/ParamAmp.txt};
			\addplot[name path=F, color = black, dashed, line width=0.95pt] table[x=Actual, y=Actual] {figures_tikz/ParamAmp.txt};
			\addplot[name path=G, color = blue,opacity=0.4] table[x=Actual, y=Up] {figures_tikz/ParamAmp.txt};			
			\tikzfillbetween[of=F and G, on layer=ft]{blue, opacity=0.25};
			\addplot[name path=G, color = blue,opacity=0.4] table[x=Actual, y=Down] {figures_tikz/ParamAmp.txt};			
			\tikzfillbetween[of=F and G, on layer=ft]{blue, opacity=0.25};
			\addlegendentry{Mean estimation}
			\addlegendentry{Perfect estimation}
			\addlegendentry{Confidence bounds}
					
		\end{axis}
		
	\end{tikzpicture}
        \caption{Parameter estimation for excitation's angle $\theta$.}\label{fig:kingpin_param_theta}
    \end{subfigure}
	\caption{Kingpin system: Performance on the parameter inference task for two example parameters with three standard deviation confidence bounds.}
	\label{fig:kingpin_paramest}
\end{figure}
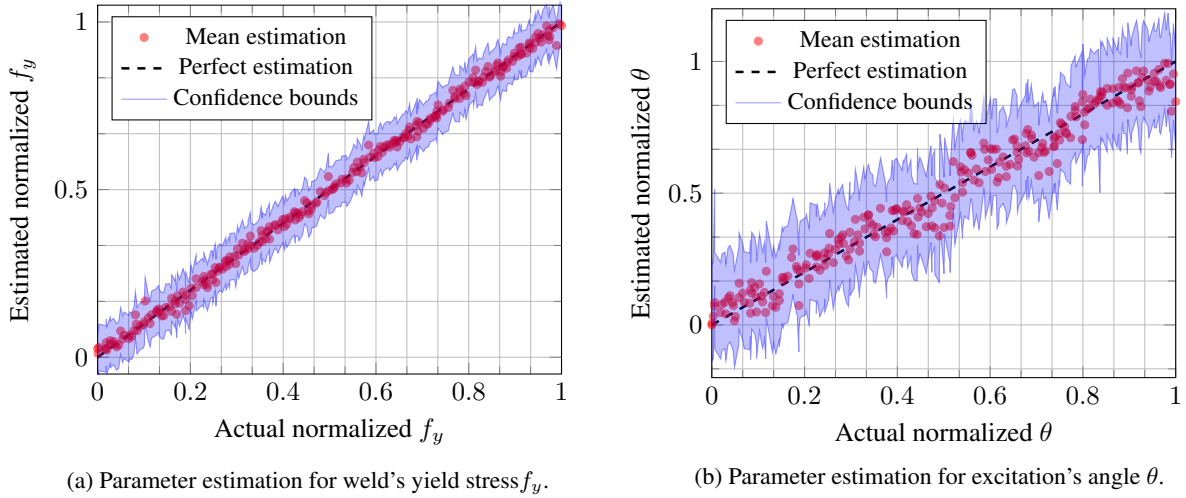

The framework's performance with respect to the parameter inference task, as formulated in \autoref{eq:param_estim}, is reported in \autoref{fig:kingpin_paramest}.
As observed in \autoref{fig:kingpin_param_yield}, the inferred normal distribution for the yield stress of the weld component captures the underlying one, producing an accurate mapping with narrow confidence bounds. 
Similar behavior is observed when estimating the yield stress of the kingpin component.

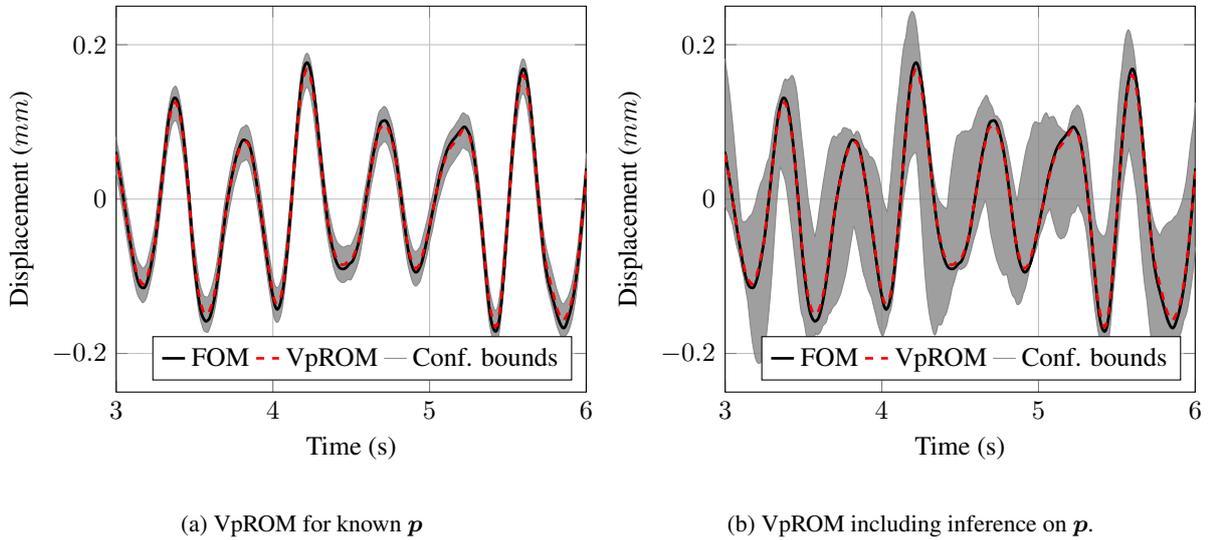
\begin{figure}[!bht]
    \begin{subfigure}[c]{0.48\textwidth}
        \pgfplotsset{
compat=1.11,
legend image code/.code={
\draw[mark repeat=2,mark phase=2]
plot coordinates {
(0cm,0cm)
(0.15cm,0cm)        
(0.3cm,0cm)         
};%
}
}

\begin{tikzpicture}[]
	\begin{axis}[
		name = response,
		xmin = 3,
		xmax = 6,
		ymin = -0.25,
		ymax = 0.25,
		xtick = {3,4,5,6},
		ytick = {-0.2,0.0,0.2},
		xlabel = {Time (s)},
		ylabel = {Displacement ($mm$)},
		grid = both,
		width=0.98\textwidth,
		legend style={legend columns=-1, legend pos=south east}
		]
		
		\addplot[name path=F, color = black, line width=1pt] table[x=Time, y=FOM] {figures_tikz/UQ_Basis_Params_Median.txt};
		\addplot[color = red, dashed, line width=1pt] table[x=Time, y=ROM]  {figures_tikz/UQ_Basis_Params_Median.txt};
		\addplot[name path=G, color = gray,opacity=0.90] table[x=Time, y=BasisUp]  {figures_tikz/UQ_Basis_Params_Median.txt};			
		\tikzfillbetween[of=F and G,on layer=main]{gray, opacity=0.75};
		\addplot[name path=G, color = gray,opacity=0.90] table[x=Time, y=BasisInf]  {figures_tikz/UQ_Basis_Params_Median.txt};		
		\tikzfillbetween[of=F and G,on layer=main]{gray, opacity=0.75};
        \addplot[name path=F, color = black, line width=1pt] table[x=Time, y=FOM] {figures_tikz/UQ_Basis_Params_Median.txt};
		\addplot[color = red, dashed, line width=1pt] table[x=Time, y=ROM]  {figures_tikz/UQ_Basis_Params_Median.txt};
		\addlegendentry{FOM}
		\addlegendentry{VpROM}
        \addlegendentry{Conf. bounds}
  
		\end{axis}
	
\end{tikzpicture}
        \caption{VpROM for known $\bp$}  \label{fig:kingpin_BasisUQ}
    \end{subfigure}     
    \begin{subfigure}[c]{0.48\textwidth}
        \pgfplotsset{
compat=1.11,
legend image code/.code={
\draw[mark repeat=2,mark phase=2]
plot coordinates {
(0cm,0cm)
(0.15cm,0cm)        
(0.3cm,0cm)         
};%
}
}

\begin{tikzpicture}[]
	\begin{axis}[
		name = response,
		xmin = 3,
		xmax = 6,
		ymin = -0.25,
		ymax = 0.25,
		xtick = {3,4,5,6},
		ytick = {-0.2,0.0,0.2},
		xlabel = {Time (s)},
		ylabel = {Displacement ($mm$)},
		grid = both,
		width=0.98\textwidth,
		legend style={legend columns=-1, legend pos=south east}
		]
		
		\addplot[name path=F, color = black, line width=1pt] table[x=Time, y=FOM] {figures_tikz/UQ_Basis_Params_Median.txt};
		\addplot[color = red, dashed, line width=1pt] table[x=Time, y=ROM]  {figures_tikz/UQ_Basis_Params_Median.txt};
		\addplot[name path=G, color = gray,opacity=0.90] table[x=Time, y=ParamUp]  {figures_tikz/UQ_Basis_Params_Median.txt};			
		\tikzfillbetween[of=F and G,on layer=main]{gray, opacity=0.75};
		\addplot[name path=G, color = gray,opacity=0.90] table[x=Time, y=ParamInf]  {figures_tikz/UQ_Basis_Params_Median.txt};		
		\tikzfillbetween[of=F and G,on layer=main]{gray, opacity=0.75};
        \addplot[name path=F, color = black, line width=1pt] table[x=Time, y=FOM] {figures_tikz/UQ_Basis_Params_Median.txt};
		\addplot[color = red, dashed, line width=1pt] table[x=Time, y=ROM]  {figures_tikz/UQ_Basis_Params_Median.txt};
		\addlegendentry{FOM}
		  \addlegendentry{VpROM}
        \addlegendentry{Conf. bounds}
  
		\end{axis}
	
\end{tikzpicture}
        \caption{VpROM including inference on $\bp$.}\label{fig:kingpin_ParamUQ}
    \end{subfigure}
	\caption{Kingpin system: Quality of the VpROM approximation with and without the parameter inference. The average performance over the samples is reported with three standard deviation confidence bounds.}
	\label{fig:kingpin_UQ}
\end{figure}

In contrast, the neural network utilized struggles to capture the exact distribution of the excitation traits, as visualized in \autoref{fig:kingpin_param_theta}. 
The respective confidence bounds are considerably larger, whereas a few outliers are also present. 
However, as already highlighted, the utility of the framework does not necessarily arise from capturing the actual response traits with high precision but rather from providing accurate response envelopes in a robust manner.
Thus, the average fit is considered sufficiently accurate, and the performance of the \emph{VpROM} is discussed next. 

The average approximation quality of the \emph{cVAE} basis generation scheme is visualized in \autoref{fig:kingpin_BasisUQ}.
The \emph{VpROM} delivers an accurate response approximation with narrow bounds when neglecting the parameter inference task and assuming that $\bp$ is known, as shown in \autoref{fig:kingpin_BasisUQ}.
This performance is similar to the one obtained for the case study of the arc structure and to previous works using the \ac{cVAE} component \cite{simpson2023vprom} and to the state-of-the-art approaches available in the literature, e.g., \cite{barnett2023neural}. 
The performance of the fully-assembled \emph{VpROM} after incorporating the parameter inference task and the \ac{cVAE} basis generation is discussed next.

\subsubsection*{\textbf{Performance of the fully-assembled hyper-reduced \emph{VpROM} framework}}

Contrary to what was observed in the first case study of the arc structure, in this example, the parameter inference task introduces a higher uncertainty in the response prediction. 
This is visualized in \autoref{fig:kingpin_ParamUQ}, where the average performance of the fully-assembled \emph{VpROM} over the testing samples is reported.
The respective confidence bounds are considerably larger compared to  \autoref{fig:kingpin_BasisUQ}, where the \emph{VpROM} delivers an accurate response approximation with narrow bounds when neglecting the parameter inference task.
This is directly related to the neural network's performance in estimating the parameters, as already discussed in \autoref{fig:kingpin_paramest}.
This discrepancy implies that the uncertainty on capturing the parameter values propagates in the time history response and emerges in the form of large response bounds.
Nonetheless, the \ac{FOM} response lies within the provided bounds at all times, and the proposed \emph{VpROM} demonstrates its ability to quantify the uncertainty involved during every step of the reduction process.
The respective measures are summarized in \autoref{tab:Kingpin_UQ}.

\begin{table}[!bth]
    \centering
	\caption{Kingpin system: Response approximation error and uncertainty during every step of the \emph{VpROM} derivation. The average performance is reported, along with the standard deviation of the average variance.}
	\label{tab:Kingpin_UQ}
		\begin{tabular} { |l|  c|  c| c| c|}
        \hline
         & Truncation & Hyper-reduction & \ac{cVAE} basis interp. & Parameter inference\\
         \hline
         Error $\epsilon_u$ & $<$ 0.1\% & 1.72\% & 4.02\% & 5.73\%\\
         Upper uncertainty bound & - & - & +0.022 (mm) & +0.092 (mm) \\
         Lower uncertainty bound & - & - & -0.024 (mm) & -0.094 (mm) \\         
	    \hline
    \end{tabular}
\end{table}

The difference in the approximation quality between the two case studies mentioned earlier might imply that for the kingpin system the dynamics do not exhibit the same variability with the parametric input as in the case of the arc structure. 
For this reason, the \ac{cVAE} can capture the dominant vector modes and reproduce the response more accurately in \autoref{fig:kingpin_BasisUQ} compared to the arc structure in \autoref{fig:arc_basis_est}.
In addition, although the \ac{cVAE} and the neural networks for the parameter inference task use the same features as input, the variability of the dynamics might be dominated mainly by one or some of the parameters.
This will cause the \ac{cVAE} to deliver a more robust and accurate performance than the neural networks that need to approximate all separate parameter values.

\begin{figure}[!hbt]
\begin{tikzpicture}[spy using outlines=
	{rectangle, magnification=2.5, anchor = center, connect spies}]
	\begin{axis}[
		name = response,
		xmin = 0,
		xmax = 10,
		ymin = -0.3,
		ymax = 0.3,
		xlabel = {Time (s)},
		ylabel = {Displacement ($mm$)},
		grid = both,
		width=0.75\textwidth,
		height=0.4\textwidth,
		legend style={legend columns=-1, legend pos=south east}
		]
		
		\addplot[name path=F, color = black, line width=.50pt] table[x=Time, y=FOM] {figures_tikz/UQ_Param_Disps_Med.txt};
		\addplot[color = red, dashed, line width=0.50pt] table[x=Time, y=ROM]  {figures_tikz/UQ_Param_Disps_Med.txt};
		\addplot[name path=G, color = gray,opacity=0.90] table[x=Time, y=Up]  {figures_tikz/UQ_Param_Disps_Med.txt};			
		\tikzfillbetween[of=F and G,on layer=main]{gray, opacity=0.75};
		\addplot[name path=G, color = gray,opacity=0.90] table[x=Time, y=Down]  {figures_tikz/UQ_Param_Disps_Med.txt};		
		\tikzfillbetween[of=F and G,on layer=main]{gray, opacity=0.75};
        \addplot[color = black, line width=0.75pt] table[x=Time, y=FOM] {figures_tikz/UQ_Param_Disps_Med.txt};
		\addplot[color = red, line width=0.75pt] table[x=Time, y=ROM]  {figures_tikz/UQ_Param_Disps_Med.txt};
		\addlegendentry{FOM}
		\addlegendentry{(Hyper-reduced) VpROM}
        \addlegendentry{Confidence Bounds}
		
		\coordinate (spypoint) at (axis cs:6.80,0.17);
		
		\end{axis}
		
		\path node[anchor= center] (magnifyglass) at (0.8\textwidth,0.15\textwidth) {};
		
		\spy [blue, every spy on node/.append style={ultra thick}, width=0.25\textwidth, height=0.30\textwidth] on (spypoint)
		in node[fill=white] at (magnifyglass);

\end{tikzpicture}
\captionsetup{skip=-5pt}
\caption{Kingpin system: Quality of the VpROM displacement response approximation. The maximum error approximation over the parametric samples is reported with three standard deviation confidence bounds.}
	\label{fig:kingpin_UmedTH}
\end{figure}
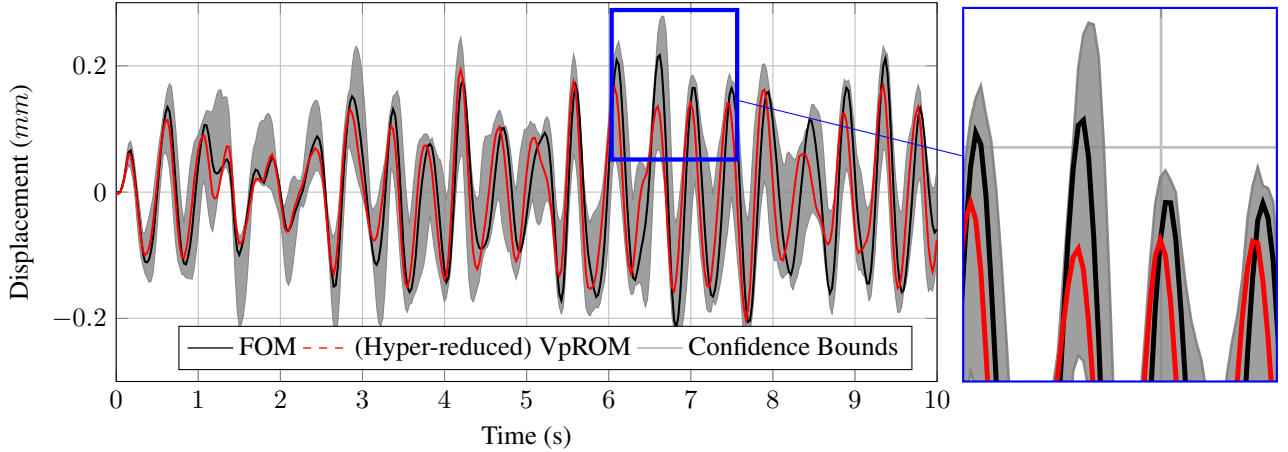

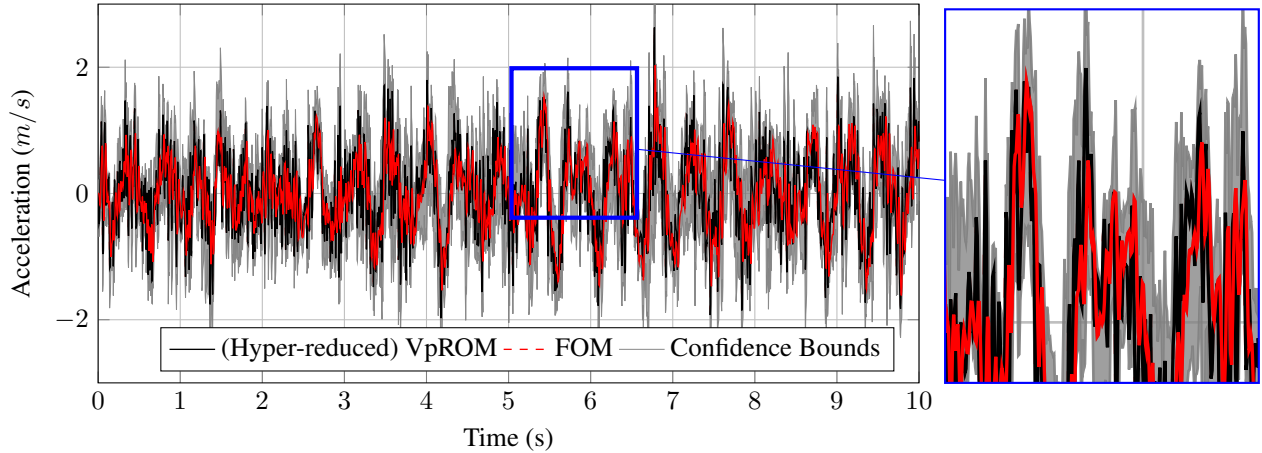
\begin{figure}[!htb]
\begin{tikzpicture}[spy using outlines=
	{rectangle, magnification=2.5, anchor = center, connect spies}]
	\begin{axis}[
		name = response,
		xmin = 0,
		xmax = 10,
		ymin = -3,
		ymax = 3,
		xlabel = {Time (s)},
		ylabel = {Acceleration ($m/s$)},
		grid = both,
		width=0.75\textwidth,
		height=0.4\textwidth,
		legend style={legend columns=-1, legend pos=south east}
		]
		
		\addplot[name path=F, color = black, line width=0.50pt] table[x=Time, y=FOM] {figures_tikz/UQ_Param_Acc_Med.txt};
		\addplot[color = red, dashed, line width=0.50pt] table[x=Time, y=ROM]  {figures_tikz/UQ_Param_Acc_Med.txt};
		\addplot[name path=G, color = gray,opacity=0.90] table[x=Time, y=Up]  {figures_tikz/UQ_Param_Acc_Med.txt};			
		\tikzfillbetween[of=F and G,on layer=main]{gray, opacity=0.75};
		\addplot[name path=G, color = gray,opacity=0.90] table[x=Time, y=Down]  {figures_tikz/UQ_Param_Acc_Med.txt};		
		\tikzfillbetween[of=F and G,on layer=main]{gray, opacity=0.75};
		\addplot[color = black, line width=0.50pt] table[x=Time, y=ROM]  {figures_tikz/UQ_Param_Acc_Med.txt};
        \addplot[color = red, line width=0.50pt] table[x=Time, y=FOM] {figures_tikz/UQ_Param_Acc_Med.txt};
		\addlegendentry{(Hyper-reduced) VpROM}
        \addlegendentry{FOM}
        \addlegendentry{Confidence Bounds}
		
		\coordinate (spypoint) at (axis cs:5.8,0.8);
		
		\end{axis}
		
		\path node[anchor= center] (magnifyglass) at (0.8\textwidth,0.15\textwidth) {};
		
		\spy [blue, every spy on node/.append style={ultra thick}, width=0.25\textwidth, height=0.30\textwidth] on (spypoint)
		in node[fill=white] at (magnifyglass);

\end{tikzpicture}
\captionsetup{skip=-5pt}
\caption{Kingpin system: Quality of the VpROM acceleration response approximation. The maximum error approximation over the parametric samples is reported with three standard deviation confidence bounds.}
	\label{fig:kingpin_AmedTH}
\end{figure}

After coupling the parameter inference neural network and the \ac{cVAE} basis generation component and equipping the resulting model with hyper-reduction, the derived hyper-reduced \emph{VpROM} can capture the dynamic response and propagate the behavior forward in time, obtaining time history estimates for all response fields of interest at once. 
This is exemplified in Figures \ref{fig:kingpin_UmedTH}, \ref{fig:kingpin_AmedTH} for the maximum error approximation for displacements and accelerations respectively.
Due to the inherent complexity of the parameter estimation problem from limited output measurements already discussed in \autoref{fig:kingpin_param_theta}, the average approximation of the hyper-reduced \emph{VpROM} does not reproduces the underlying \ac{FOM} response with high precision, as depicted in \autoref{fig:kingpin_UmedTH}. 
An example testing sample where the \emph{VpROM} delivers the maximum error approximation is depicted, however, a similar observation can be made for the response approximation over the rest of the testing samples. 
Although the response agreement between the \ac{FOM} and the derived virtual representation is crucial for the optimal solution of \ac{SHM} problems and decision-making, the latter can be still addressed in a sub-optimal sense via quantifying the involved uncertainties.
The inherent probabilistic nature of the framework's basis generation and parameter estimation components achieves this, thus resulting to a time history approximation with response bounds that function as response envelopes. 

Concretely, by sampling the latent space of the \ac{cVAE} using one hundred additional draws and utilizing the respective distribution captured by the parameter estimation neural network $\Tilde{N}$, one hundred \emph{VpROM} runs are evaluated in parallel.
The resulting approximations are then aggregated in the form of envelope curves that build the confidence bounds illustrated by the shaded regions in all time history approximations. 
The corresponding bounds contain the actual \ac{FOM} response even in themaximum error approximation, visualized in \autoref{fig:kingpin_UmedTH} for the displacement time history and in \autoref{fig:kingpin_AmedTH} for the accelerations, indicating a robust \emph{VpROM} able to provide accurate uncertainty bounds over an operational parametric range based on limited incoming measurements.

Lastly, the computational efficiency measures are summarized in \autoref{tab:Kingpin_efficiency}.
During training, all \ac{FOM} model evaluations are carried out in parallel to minimize the necessary resources.
Regarding the training process of the \ac{cVAE} basis generation component, the hyper-reduced \emph{VpROM} is evaluated during the later stages of the training process when the \emph{VpROM}'s accuracy term of the loss function in \autoref{eq:lossA} is activated. 
Details on the respective setup are reported in \autoref{sec:app2}.
The proposed framework achieves an accelerated model evaluation, with a speedup factor of 24.89 compared to the \ac{FOM} simulation. 
With respect to the toll reduction, the bases of the hyper-reduced \emph{VpROM} retain only eight components, compared to the 126621 \ac{DOF}s of the \ac{FOM} model, whereas only 3317 elements are retained for the respective nonlinear terms evaluation, indicating an approximately 95\% reduction.

\begin{table}[!hbt]
    \centering
	\caption{Kingpin system: Computational efficiency of the \emph{VpROM} and its hyper-reduced version. Timings exclude the training phase and refer to online model evaluations.}
	\label{tab:Kingpin_efficiency}
		\begin{tabular} { |l|  c|  c| c|}
        \hline
         & \ac{FOM} & \emph{VpROM} & Hyper-reduced \emph{VpROM}\\
         \hline
         model size             &    126621       &      8     &      8     \\
         elements                &    52044        &    52044  &      3317       \\ 
         Training time (s)       &    -       &    84600   &      84600\\
         Solution time (s)       &    15661       &    1478   &      638\\
         Speedup ($t_{FOM}/t_{ROM}$)                  &    1.00       &    10.42   &      24.89   \\
	    \hline
    \end{tabular}
\end{table}

\section{Limitations and concluding remarks}
\label{sec:Conclusions}

This work demonstrates the use of a \ac{cVAE}-based \ac{ROM}, termed as \emph{VpROM}, for response prediction in parametric nonlinear systems. 
The main contribution of this work lies in conditioning the local reduction bases to monitoring features that can be obtained online while also mapping the physical parameter space of the \ac{ROM} to the same features to recover the parameter values needed to propagate the dynamics in the low-order space. 
This is different from the practice appearing in literature so far, which relies on the prescription of the physics-based system parameters for the computation of the respective bases. 
While this approach is efficient for forward problem formulations, it becomes impractical within an inverse context, such as the one pertaining to the \ac{SHM} case.
The training process of the \ac{cVAE} basis generation component is refined compared to prior efforts in \cite{simpson2023vprom} for increasing robustness. 

The following observations are made:
\begin{itemize}
    \item We show that \ac{cVAE} can be used to generate efficient local reduction bases for nonlinear parametric \ac{ROM}s, while neural networks to recover the link between monitoring features and physical parameters via a parametric formulation of suitable probability distributions.
    The conditioning on monitoring features renders the scheme practical for use with data that are derived from sensors, feasibly in an online manner (near real-time).    
    \item This naturally comes with the challenge of inferring physical quantities from often limited output measurements. When the sensing grid is sparse, the proposed \emph{VpROM} may struggle to capture the \ac{FOM} response with high precision; however, our formulation offers the additional benefit of encoding the uncertainty on the delivered prediction. Via quantifying the respective error bounds during every step of the reduction and prediction process, the proposed \ac{ROM} delivers response envelopes that capture the underlying dynamic behavior robustly in all response fields.
    \item The demonstration of the proposed framework on a large-scale numerical case study results in substantial toll reduction and a significantly accelerated \ac{ROM}, almost 25 times faster than the \ac{FOM}.
\end{itemize}

The approach proposed in this study is demonstrated on two simulated nonlinear systems, parameterized in terms of the system properties and the loading traits. 
The first example employs an arc structure exhibiting geometric nonlinearity due to large deformations.
The second case study demonstrates the ability of the \emph{VpROM} to address large-scale systems and constitutive material nonlinearities. 
In both examples, the potential of the framework for quantifying uncertainty and providing confidence bounds on the obtained estimates is also reported.

Compared to state-of-the-art methodologies, the relative complexity of the training process of the \emph{VpROM} and the respective computational budget needed may pose a limitation. 
Certain heuristics, such as grid search, exist for selecting optimal hyper-parameters, including number of layers, neurons per layer, activation functions, or learning rates, yet the training process requires substantially increased effort compared to current clustering or interpolation methods.
The Gaussian nature of the latent space of the \ac{cVAE}, along with the Gaussian probability assumption imposed by the neural networks that are tasked with inferring the physical parameters from the monitored features, pose additional limitations and serve as points to be taken on in future extensions of the \emph{VpROM} scheme.

Furthermore, the proposed framework has been exemplified in two numerical examples featuring smoothly varying nonlinear behavior. 
Although both computational plasticity and geometric nonlinearities due to large deformations have been studied, indicating a sufficient variety of nonlinear effects, their variability with respect to the parametric samples is or assumed to be continuous and relatively smooth. 
However, case studies featuring contact nonlinearities or response domains with discontinuities like the ones caused by cracks require dedicated treatment.
Such examples are common in \ac{SHM} and necessitate further extensions of the proposed framework. 
The same applies to chaotic systems or complex response bifurcations that were not examined here.

\section*{Data Availability }
The data supporting this study's findings are available from the author \textbf{KV} upon reasonable request.

\section*{Declaration of Competing Interest}
The authors declare that they have no known competing financial interests or personal relationships that could have appeared to influence the work reported in this paper.

\section*{Acknowledgements}

The authors gratefully acknowledge the funding from the European Commission under the Horizon Europe funding guarantee, for the project ‘ReCharged - Climate-aware Resilience for Sustainable Critical and interdependent Infrastructure Systems enhanced by emerging Digital Technologies’ (grant agreement No: 101086413) and from the Sandia National Laboratories. Sandia National Laboratories is a multimission laboratory managed and operated by National Technology and Engineering Solutions of Sandia, LLC, a wholly owned subsidiary of Honeywell International Inc., for the U.S. Department of Energy’s National Nuclear Security Administration under contract DE-NA0003525.  This paper describes objective technical results and analysis. Any subjective views or opinions that might be expressed in the paper do not necessarily represent the views of the U.S. Department of Energy or the United States Government.

\bibliographystyle{plain}
\bibliography{main}

\section*{CRediT authorship contribution statement}
\textbf{Konstantinos Vlachas}: Conceptualization, Methodology, Software, Validation, Writing.
\textbf{Thomas Simpson}: Software. 
\textbf{Anthony Garland}: Methodology, Resources, Review \& editing.
\textbf{D. Dane Quinn}: Methodology, Review \& editing.
\textbf{Charbel Farhat}: Conceptualization, Methodology, Review \& editing, Supervision.
\textbf{Eleni Chatzi}: Conceptualization,Methodology, Resources, Writing - review \& editing, Supervision.

\clearpage
\newpage
\renewcommand{\thefigure}{A\arabic{figure}}
\setcounter{table}{0}
\appendix 
\section{Order selection of POD bases} \label{sec:app2}

The reduced order $r$ of the \ac{POD} projection bases is related to the projection error of \autoref{eq:project}.
Specifically, it has been shown that the sum of the squares of the neglected singular values $\sigma$ due to truncation in \autoref{eq:Vmodes} scales with the projection error in \autoref{eq:project} \cite{volkwein2013proper}.  
Thus, following the respective suggestions in the available literature, the following heuristic criterion is used to select the truncation order $r$:
\begin{equation}
    \frac{\sum\limits_{i=r+1}^{n} \sigma_i^2}{\sum\limits_{i=1}^{n}\sigma_i^2} \leq \epsilon
\end{equation}
where $\epsilon \in [0, 1]$ is a user-defined tolerance parameter.
As $\epsilon$ tends to zero, the retained modes $\bL_i$ increase, thus increasing the truncation order and yielding a more accurate representation with respect to the projection in \autoref{eq:project}.
However, in many systems, especially structural engineering ones, the singular values typically exhibit an initial fast decay, followed by a considerably slower reduction rate, as demonstrated in \cite{Vlachas2021}.
Thus, a proper reduced order that retains the dominant modes can be selected by inspecting the decay rate on the respective graph of the singular values $\sigma$.
In the numerical examples of this study, this is achieved by setting $\epsilon=1e^{-5}$.

\section{Network Architectures used} \label{sec:app1}

\subsection{Arc structure with large deformations}

The architecture of the \ac{cVAE} network used for the arc case study along with the hyper-parameter values and the details of the training process are summarized in \autoref{tab:arc_VAE}.
The respective architecture for the neural network utilized for parameter inference is reported in \autoref{tab:arc_NN}, along with the necessary details defining its training.

\begin{table}[!htb]
\begin{minipage}{.5\linewidth}
    \centering
    \medskip	
	\begin{tabular} { |c|  c|  c| c|}
        \hline
          & Type & Weights & Activation\\
         \hline
         \multirow{4}{*}{Encoder} & Dense  & (1024, 64) & \textit{tanh} \\
         & Dense  & (64,64)   & \textit{tanh} \\ 
         & Dense  &  (64,64)  & \textit{tanh}    \\
         & Dense  &  (64,12)  & \textit{linear}   \\ 
         \hline
         \multirow{4}{*}{Decoder}& Dense  & (12, 64) & \textit{linear} \\
         & Dense  & (64,64)   & \textit{tanh} \\ 
         & Dense  &  (64,64)  & \textit{tanh}    \\
         & Dense  &  (64,1024)  & \textit{linear}   \\ 
	    \hline
    \end{tabular}
\end{minipage}\hfill
\begin{minipage}{.5\linewidth}
    \centering
    \medskip	
	\begin{tabular} { |c|  c|}
        \hline
         Hyper-parameter & Value \\
         \hline
         Optimizer & Adam   \\
         Initial learning rate  & 0.0001   \\ 
         Max epochs & 1250  \\ 
         Scaling & [0,1]\\
         PCA & $\bW \in \mathrm{R}^{35\times1}$\\
         Batch size & 64\\
         $\gamma_1$,$\gamma_2$ & 0.15, 0.15 \\
         $\Tilde{r}, r$& 64, 16 \\         
	    \hline
    \end{tabular}
\end{minipage}\hfill
\caption{Arc structure: Architecture of the cVAE used for basis interpolation, along with training details.}
\label{tab:arc_VAE}
\end{table}

\begin{table}[!htb]
\begin{minipage}{.5\linewidth}
    \centering
    \medskip	
	\begin{tabular} { |c|  c|  c| c|}
        \hline
          & Type & Weights & Activation\\
         \hline
         \multirow{4}{*}{Mean layer} & Dense (shared)  & (35, 256) & \textit{ReLU} \\
         & Dense  & (256, 64)   & \textit{ReLU} \\ 
         & Dense  & (64, 16)   & \textit{ReLU} \\ 
         & Dense  &  (16, 1)  & \textit{linear}    \\
         \hline
         \multirow{4}{*}{Std layer}& Dense (shared)  & (35, 256) & \textit{ReLU} \\
         & Dense  & (256, 64)   & \textit{ReLU} \\ 
         & Dense  & (64,16)   & \textit{ReLU} \\ 
         & Dense  &  (16,1)  & \textit{Softplus}   \\ 
	    \hline
    \end{tabular}
\end{minipage}\hfill
\begin{minipage}{.5\linewidth}
    \centering
    \medskip	
	\begin{tabular} { |c|  c|}
        \hline
         Hyper-parameter & Value \\
         \hline
         Optimizer & Adam   \\
         Initial learning rate  & 0.001   \\ 
         Max epochs & 1000  \\ 
         Scaling & [0,1]\\
         PCA & $\bW \in \mathrm{R}^{35\times1}$\\
         Cross-validation & 5 folds\\   
	    \hline
    \end{tabular}
\end{minipage}\hfill
\caption{Arc structure: Architecture of the neural network used for parameter inference, along with training details.}
\label{tab:arc_NN}
\end{table}

\subsection{Kingpin system with plasticity} 

The architecture of the \ac{cVAE} network used for the kingpin system case study along with the hyper-parameter values and the details of the training process are summarized in \autoref{tab:kingpin_VAE}.
The respective architecture for the neural network utilized for parameter inference is similar to the one reported for the arc structure in \autoref{tab:arc_NN}, with the input dimension adjusted according to the kingpin system.
The \ac{NARX} model is implemented using the \emph{MATLAB} neural net time series application and the respective details are summarized in \autoref{tab:NARX}.
All \ac{NARX} models have the same training properties and architecture.

\begin{table}[!htb]
\begin{minipage}{.50\linewidth}
    \centering
    \medskip	
	\begin{tabular} { |c|  c|  c| c|}
        \hline
          & Type & Weights & Activation\\
         \hline
         \multirow{4}{*}{Encoder} & Dense  & (512, 64) & \textit{tanh} \\
         & Dense  & (64,64)   & \textit{tanh} \\ 
         & Dense  &  (64,64)  & \textit{tanh}    \\
         & Dense  &  (64,12)  & \textit{linear}   \\ 
         \hline
         \multirow{4}{*}{Decoder}& Dense  & (12, 64) & \textit{linear} \\
         & Dense  & (64,64)   & \textit{tanh} \\ 
         & Dense  &  (64,64)  & \textit{tanh}    \\
         & Dense  &  (64,512)  & \textit{linear}   \\ 
	    \hline
    \end{tabular}
\end{minipage} \hfill
\begin{minipage}{.50\linewidth}
    \centering
    \medskip	
	\begin{tabular} { |c|  c|}
        \hline
         Hyper-parameter & Value \\
         \hline
         Optimizer & Adam   \\
         Initial learning rate  & 0.0001   \\ 
         Max epochs & 1250  \\ 
         Scaling & [0,1]\\
         PCA & $\bW \in \mathrm{R}^{48\times1}$\\
         Batch size & 64\\
         $\gamma_1$,$\gamma_2$ & 0.15, 0.15 \\
         $\Tilde{r}, r$& 64, 8 \\         
	    \hline
    \end{tabular}
\end{minipage} 
\caption{Kingpin system: Architecture of the cVAE used for basis interpolation, along with hyper-parameter values.}
\label{tab:kingpin_VAE}
\end{table}

\begin{table}[!htbp]
\caption{Kingpin system: Architecture of the NARX used for feature extraction.}
\label{tab:NARX}
\centering
\begin{tabular}{l c c c c c c}
\hline 
Parameter: & Time delay & Layer size & Training method & Epochs & Activation & Output layer \\
Value: & 2 & 10  & Bayesian regularization & 1000 & Tan-Sigmoid & Linear \\
\hline
\end{tabular}
\end{table}

\end{document}